\documentstyle[12pt,epsfig]{article}

\newlength{\dinwidth}
\newlength{\dinmargin}
\begin{document}
\def\bold#1{\setbox0=\hbox{$#1$}%
     \kern-.025em\copy0\kern-\wd0
     \kern.05em\%\baselineskip=18ptemptcopy0\kern-\wd0
     \kern-.025em\raise.0433em\box0 }
\def\slash#1{\setbox0=\hbox{$#1$}#1\hskip-\wd0\dimen0=5pt\advance
         to\wd0{\hss\sl/\/\hss}}
\newcommand{\be}{\begin{equation}}
\newcommand{\ee}{\end{equation}}
\newcommand{\bea}{\begin{eqnarray}}
\newcommand{\eea}{\end{eqnarray}}
\newcommand{\nn}{\nonumber}
\newcommand{\dd}{\displaystyle}
\newcommand{\bra}[1]{\left\langle #1 \right|}
\newcommand{\ket}[1]{\left| #1 \right\rangle}
\newcommand{\spur}[1]{\not\! #1 \,}
\thispagestyle{empty} \vspace*{1cm}
\rightline{BARI-TH/04-486}
\vspace*{2cm}
\begin{center}
  \begin{LARGE}
  \  \begin{bf}
Excited  Charmed  Mesons:\\
\vspace*{0.3cm}
Observations, Analyses\\
\vspace*{0.4cm}
 and Puzzles
\vspace*{0.5cm}
 \end{bf}
  \end{LARGE}
\end{center}
\vspace*{8mm}
\begin{center}
\begin{large}
P.  Colangelo, F. De Fazio and R. Ferrandes
\end{large}
\end{center}
\begin{center}
\begin{it}
Istituto Nazionale di Fisica Nucleare, Sezione di Bari, Italy
\end{it}
\end{center}
\begin{quotation}
\vspace*{1.5cm}
\begin{center}
  \begin{bf}
  Abstract\\
  \end{bf}
\end{center}
\vspace*{0.5cm} \noindent We review the status of  recently
observed positive parity charmed resonances, both in the
non-strange and in the strange sector. We describe the
experimental findings, the main theoretical analyses and the open
problems deserving further investigations.
\vspace*{0.5cm}
\end{quotation}
\newpage
\baselineskip=18pt \vspace{2cm} \noindent

\section{Introduction}
This is  an exciting period for hadron spectroscopy, due to the
discovery of several new particles with unexpected and intreaguing
features.  It  is fair  to mention first the  increasing evidence
of  pentaquark states \cite{zhu},  the observation of which
requires a deeper understanding of QCD interactions at low energy.
Furthermore, both the $\bar b b$ and $\bar c c$ spectra were
enriched by the observation of a meson belonging to the $d$-wave
multiplet of the $\Upsilon$ system,  and
 of  $X(3870)$
and the first radial excitation of  $\eta_c$   in the charmonium
\cite{Skwarnicki:2003wn}. A new doubly charmed $\Xi^+_{cc}$ baryon
was detected  \cite{Xcc}.  Last, but not least, excited charmed
mesons have been observed  both in the strange and non-strange
sector,  providing new information about  the spectroscopy of the
open charm system. In this paper we  describe   the experimental
results concerning   such  new charmed resonances,  as well as  a
number of  theoretical analyses  aimed at  understanding their
phenomenology.

As we shall see, there are various and different interpretations of
these  charmed resonances, and in the following we   discuss them in detail.
However, it is important  at the beginning  to  settle the scene, and
we consider  the heavy quark theory as
the most suitable  theoretical framework to  start  our study
\cite{rassegne_hqet}.

The  analysis of hadrons containing a single heavy quark $Q=c,b$
is greatly simplified if one considers the limit of infinitely
heavy quark. This is due to the fact that  such a quark acts as a
static colour source, and its spin $s_Q$ is decoupled from the
total angular momentum $s_\ell$ of all other hadronic (light)
degrees of freedom.

There are several consequences of that.  One is that it is possible to
classify heavy hadrons  using  $s_\ell$ as a good quantum number.
Therefore, heavy mesons can be collected in doublets, each one
corresponding to a particular value of $s_\ell$ and parity,  with the
members of each doublet  degenerate in mass.
The  degeneracy condition  is broken  if  $1/m_Q $ corrections are taken into account.
In this case,   the mass formula  for a  heavy  meson:
\be
M_H=m_Q+{\bar \Lambda}+{\mu_\pi^2 -\mu_G^2 \over 2 m_Q} \label{massformula}
\ee
involves the binding energy  parameter $\bar \Lambda$   and   two parameters
  $\mu^2_\pi$ and $\mu_G^2$  representing  the matrix elements of the
kinetic energy  and the chromomagnetic operators  over the considered meson.
Such operators appear at
order ${1 \over m_Q}$ in the effective Lagrangian of the heavy
quark effective theory \cite{rassegne_hqet}.  $\mu^2_G$ depends on
the spin $J$ of the hadron and is therefore responsible of the mass
splitting between the two members of a doublet, which is a $1/m_Q$  effect.
It can be written
as: $\mu_G^2=-2\left[J(J+1)-\displaystyle{3 \over
2}\right]\lambda_2$. $\bar \Lambda$,  together with  $\lambda_2$ (or $\mu_G^2$)
and  $\mu_\pi^2$, are
independent of  the heavy quark mass and,  in $SU(3)_F$ limit,
of the flavour of the light quark.  All the three parameters,  however,
are different for different doublets.

The lowest lying $Q \bar q$ mesons correspond to
$\ell=0$ (the $s$-wave states of the quark  model) with
$ s_\ell^P={1 \over 2}^-$. This  doublet comprises  two
states with spin-parity $J^P=(0^-,1^-)$. For $\ell=1$ ($p$-wave
states of the quark model), it could be either $ s_\ell^P={1 \over 2}^+$
or $ s_\ell^P={3 \over 2}^+$.  The  two corresponding doublets
have   $J^P=(0^+,1^+)$ and $J^P=(1^+,2^+)$.
To fix the notations,  we denote
the members of the  $J^P=(0^+,1^+)$ charm doublet with
$s_\ell^P={1\over2}^+$
as $(D^*_{0},D_{1}^\prime)$
 and $(D^*_{s0},D_{s1}^\prime)$ for non-strange and strange states, respectively,
 and those of the
$J^P=(1^+,2^+)$  doublet with
$s_\ell^P={3\over2}^+$ as $(D_{1},D^*_{2})$ and
$(D_{s1},D^*_{s2})$.
Finite heavy quark mass effects  can
induce a mixing between the two axial vectors,  giving rise to
two $1^+$ mass eigenstates;  however, in the following we neglect the mixing since, as
we argue below, it can be at most of  few degrees in the case of charm.

A distinctive feature between the two $\ell=1$ doublets is their
expected width. In fact, the strong decays of the members of the $
s_\ell^P={3\over 2}^+$ doublet can proceed by emitting light
pseudoscalar mesons in $d$-wave,  while the emission is in
$s-$wave  for  the doublet  $s_\ell^P={1\over 2}^+$. Thus,
$s_\ell^P={3 \over 2}^+$  mesons are expected to be narrower  than
$s_\ell^P={1 \over 2}^+$ ones, simply due to the different
dependence of the two-body decay rates on the three-momentum of
the emitted meson.  The six  $s_\ell^P={3 \over 2}^+$   $c{\bar
u}$, $c{\bar d}$ and  $c{\bar s}$ states have been observed with
precision  at the level of few MeV for the mass and the width,
thanks, in particular,  to their narrowness and to   their
abundant production in various experimental setup,  at fixed
target, in $e^+ \, e^-$ continuum production, in $B$ and $Z^0$
decays \cite{PDG}.

The case is different for   the
members of   the doublet $s_\ell^P={1 \over 2}^+$, which are the
subject of this review paper. There have been recent
experimental observations of particles that can be recognized as
members of this  doublet in the case of charm.
However, for non-strange particles
not all the charge configurations have been observed, so far,
and there is some disagreement in the mass measurements
made by different experiments.
As  for $c \bar s$ mesons,  evidence has been
recently collected  of very narrow states, in contradiction with the expectation
of particles having broad width, a   feature  which has
prompted an intense activity   to  clarify  the issue.

Moreover,   in case of the assignment of the
newly observed states to the low-lying positive parity
doublet, the comparison of the features of
corresponding non-strange and strange mesons
arises some questions, for example concerning
 the mass difference between strange and non-strange mesons and the spin splitting
within each doublet.

In the following we review  (in Section \ref{sec:exp})  the  experimental
observations for both non-strange and strange charmed mesons,
in particular  the measurement of masses and decay branching fractions.
Then  we discuss  various theoretical studies  aimed at
 shedding  light on the structure of these mesons. In particular, in Section \ref{sec:masses}
 we consider the analyses of the spectroscopy of the newly observed states, with different
 approaches and interpretations. In Section \ref{sec:decays} we analyse the decays
 of the new charmed resonances, as  they can be useful for understanding their structure.
After a discussion concerning  the beauty sector, we present  our
conclusions  in the last Section.

\section{Observations} \label{sec:exp}

\subsection{Evidence of broad $c\overline{u}$,
$c\overline{d}$ states: $D_0^{*0}$, $D_0^{*+}$ and  $D_1^{\prime0}$ }

The first evidence of $c \bar q$ broad states  was provided by Cleo
Collaboration \cite{cleobroad}, which observed a  state of  mass  $2460$ MeV
and width $290$ MeV with  the features of  an axial vector meson.
More recently,  Belle and Focus Collaborations, looking at $D \pi$ and $D \pi \pi$
invariant mass distributions, provided further elements in support of the existence
of one scalar and one axial charmed meson that could be interpreted as the states belonging to the
 $\frac{1}{2}^{+}$ $c\overline{u}$,  $c\overline{d}$  doublets.

\begin{figure}[t]
 \begin{center}
  \includegraphics[width=0.32\textwidth] {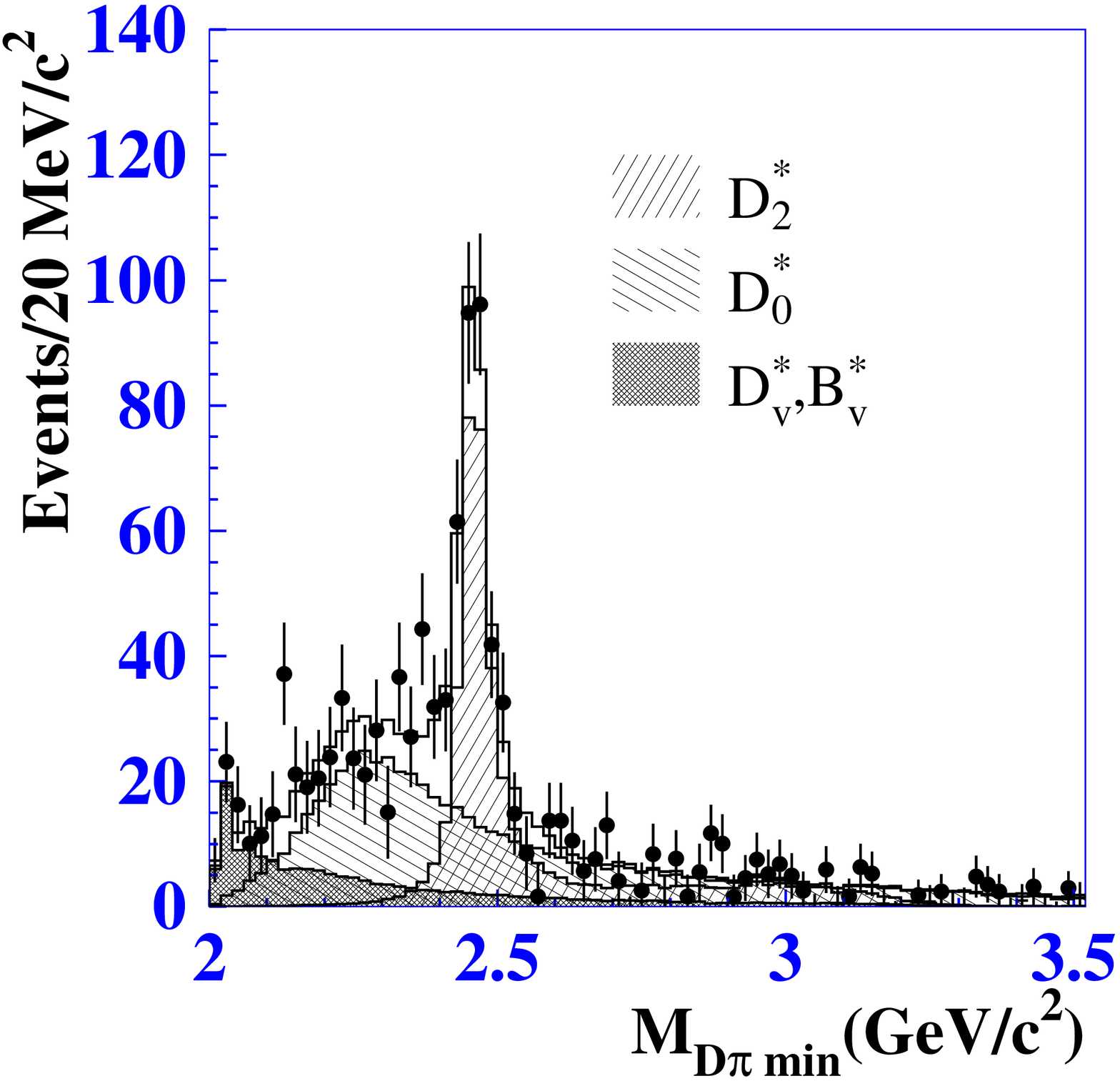}\hspace{1cm}
  \includegraphics[width=0.32\textwidth] {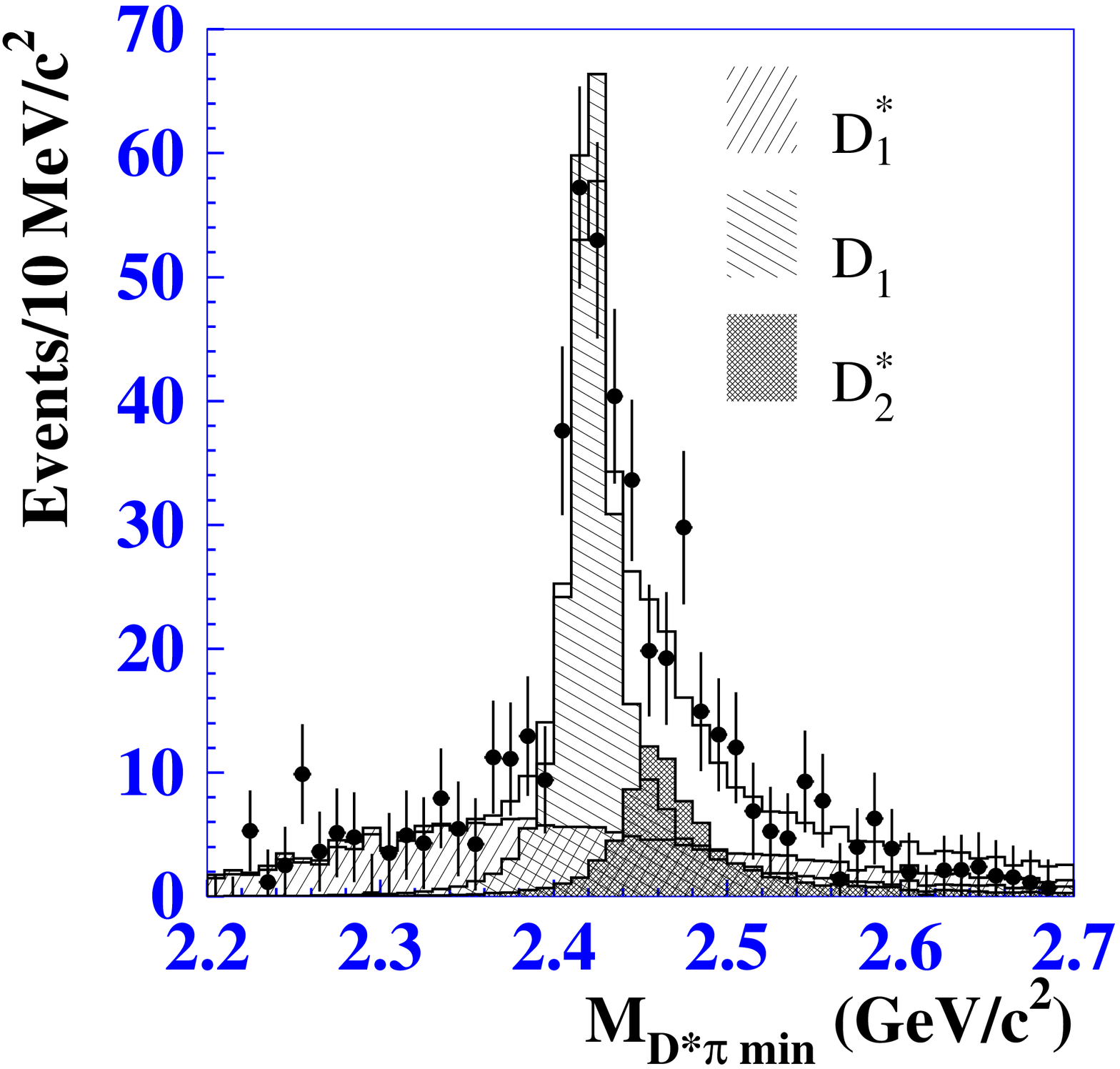}
\vspace*{0mm}
 \caption{Background-subtracted  $D\pi$ (left) and $D^*\pi$ (right)
mass distributions obtained by Belle Collaboration ~\cite{Bellelarghi}. Hatched
histograms show the contributions  of the various amplitudes, open histograms show
the coherent sum of all contributions.}
  \label{fig:Bellelarghi}
 \end{center}
\end{figure}

The Belle observation is based on the study of
charmed mesons  produced in $B$ decays  through the transition
 B$\rightarrow D^{**}\pi$, with a D$^{**}$  a generic
$l=1$ meson \cite{Bellelarghi}.
A Dalitz plot analysis is carried out for the final states
$D^+\pi^-\pi^-$ and $D^{*+}\pi^-\pi^-$. In the former case, the Dalitz
analysis includes the amplitude of  the $D_2^{*0}\pi^-$
mode, the contributions of processes involving virtual production of
$D^{*0}\pi^-$ and $B^{*0}\pi^-$, and the amplitude of a
$D \pi$  structure with free mass, free  width and
assigned $J^P=0^+$ quantum numbers.  As reported by Belle \cite{Bellelarghi},
a fit of the projection of the Dalitz plot to the $D\pi$ axis, where   the pion
is  the one having the smallest momentum,
favours the presence  of the scalar contribution. The $D \pi$ mass distribution
is depicted in fig.\ref{fig:Bellelarghi};  the mass and width of the broad state obtained
by  the fit are collected in  Table~\ref{tab:statilarghi}.

A similar analysis,  carried out  for $D^{*+}\pi^-\pi^-$, provides
evidence of a broad resonance with quantum numbers compatible with
the assignment J$^P=1^+$. The   Belle $D^*\pi$ mass distribution
is also depicted in fig.~\ref{fig:Bellelarghi}. The contributions
of the two other charmed states $D_1$ and $D^*_2$ are not
sufficient to fit the mass distribution; a further contribution,
representing a broad state, is needed, the mass and width of which
are   reported in Table~\ref{tab:statilarghi}. Through this
analysis, a new determination of mass and width of the two other
positive parity charmed states $D^*_2$ and $D_1$ has also been
obtained, together with a measurement of the mixing
angle between the two $1^+$ states: $\omega=-0.10 \pm 0.03 \pm
0.02 \pm 0.02~rad$ ($\theta\simeq -6^0$) suggesting  that such a mixing can
safely be neglected.

An analogous study has been carried out by
Focus Collaboration \cite{focus}, which considered both the $D^0 \pi^+$ and $D^+\pi^-$ charge
configurations.
 Also in this case, a broad scalar contribution
is required to fit the $D \pi$  mass distribution.  The values of
mass and width quoted by Focus are collected  in
Table~\ref{tab:statilarghi}. The  values for the mass of
$D_0^{*0}$, measured by Belle and Focus, are marginally
compatible;  nevertheless,  we include in
Table~\ref{tab:statilarghi} their  average, as well as the average of Belle and Cleo data for
the axial vector state.

\begin{table}[h]
    \caption{Mass and width  of broad resonances observed in D$\pi$ and
D$^*\pi$ systems.}
    \label{tab:statilarghi}
    \begin{center}
    \begin{tabular}{ccccc}
      \hline
      \ & \ &  Belle Collab.~\cite{Bellelarghi} &   Focus Collab.~\cite{focus}  & Average\\
      \hline
      \ $D_0^{*0}$ & \begin{tabular}{c} M (MeV) \\ $\Gamma$ (MeV)
\end{tabular}   &
      \begin{tabular}{c} $2308\pm17\pm15\pm28$ \\ $276\pm21\pm18\pm60$
\end{tabular}   &
      \begin{tabular}{c}  $2407\pm21\pm35$ \\ $240\pm55\pm59$ \end{tabular} &
      \begin{tabular}{c}  $2351\pm27$ \\ $262\pm51$ \end{tabular}
\\ \hline
         \ $D_0^{*+}$ & \begin{tabular}{c} M (MeV) \\ $\Gamma$ (MeV)
\end{tabular}   &
      \begin{tabular}{c} $$ \\ $$
\end{tabular}   &
      \begin{tabular}{c}  $2403\pm14\pm35$ \\ $283\pm24\pm34$ \end{tabular} &
      \begin{tabular}{c}  $$ \\ $$ \end{tabular}
\\

      \hline \hline
      \ & \ &  Belle Collab.~\cite{Bellelarghi} &   Cleo Collab.~\cite{cleobroad}  & Average\\
      \hline
      \ $D_1'^{0}$ & \begin{tabular}{c} M (MeV) \\ $\Gamma$ (MeV)
\end{tabular}   &
      \begin{tabular}{c} $2427\pm26\pm20\pm15$\\
                                       $384_{-75}^{+107}\pm24\pm70$\end{tabular}
      &\begin{tabular}{c} $2461^{+41}_{-34}\pm10\pm32$\\
                                       $290_{-79}^{+101}\pm26\pm36$\end{tabular}
      &\begin{tabular}{c} $2438\pm30$\\
                                       $329\pm84$\end{tabular}
                                       \\
      \hline
    \end{tabular}
  \end{center}
\end{table}

\subsection{The meson $D_{sJ}^*(2317)$}
\begin{figure}
 \begin{center}
  \includegraphics[width=0.31\textwidth] {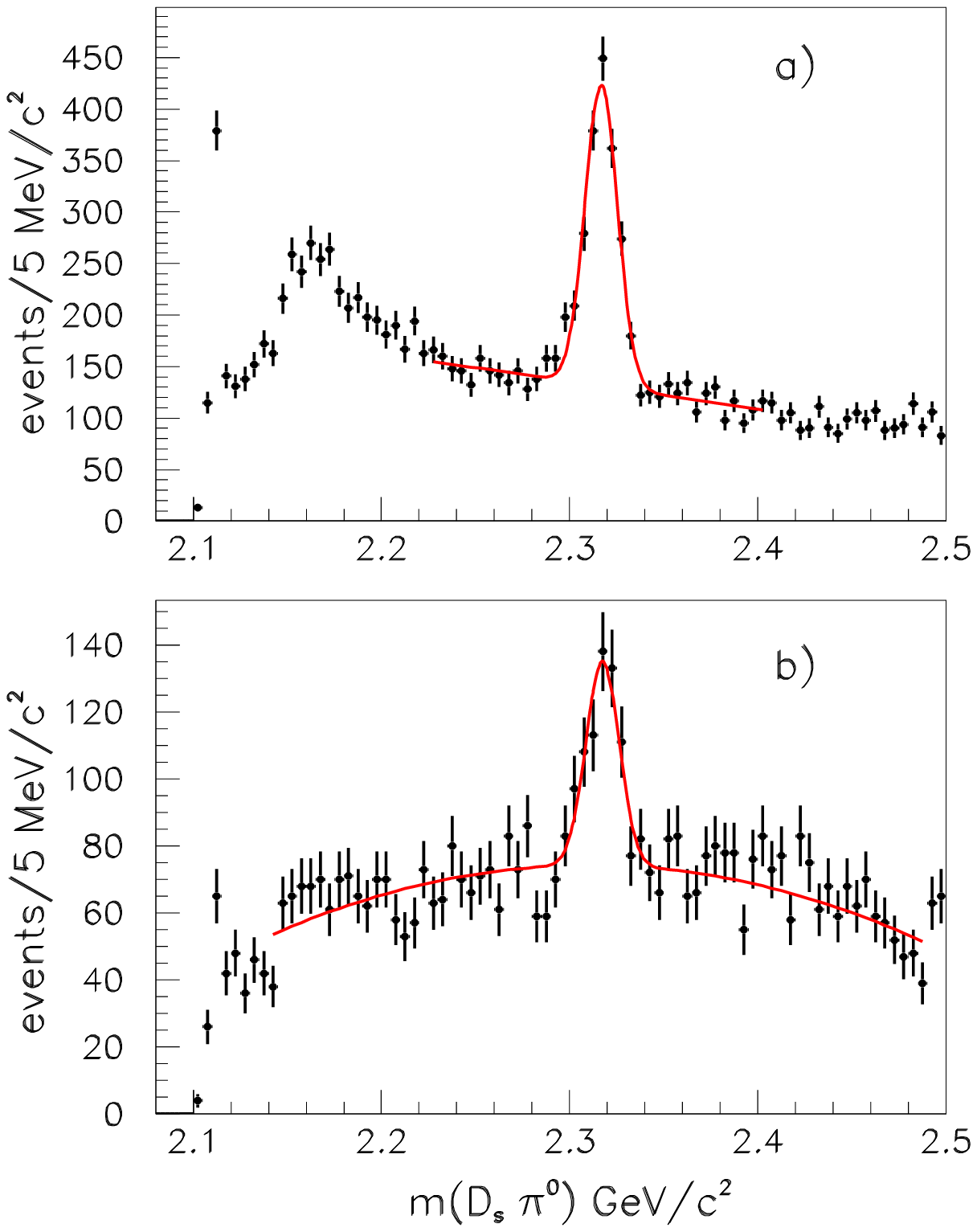}\hspace{0.7cm}
  \includegraphics[width=0.342\textwidth] {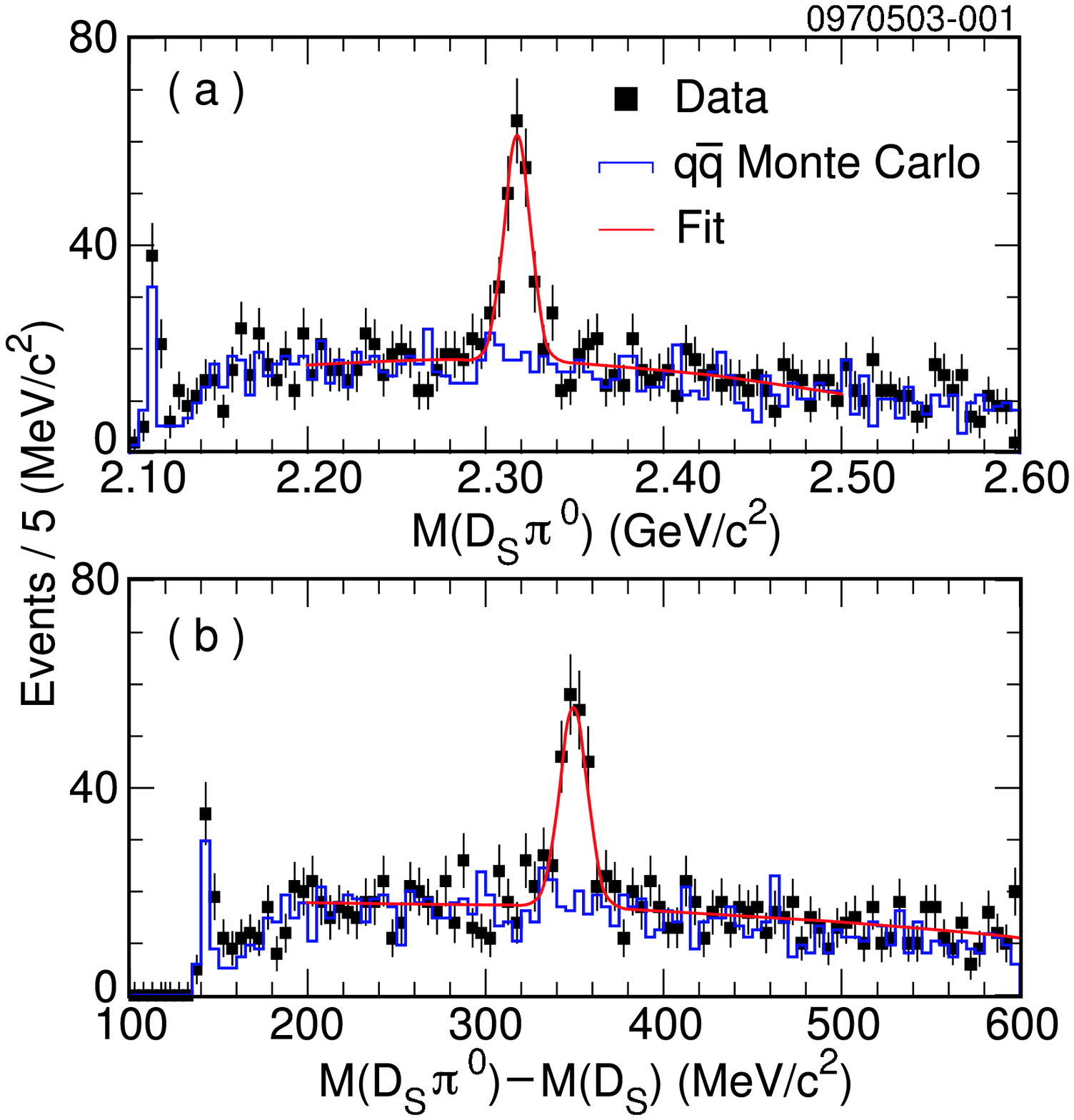}
\vspace*{0mm}
 \caption{Left:  $D_{s}^+\pi^0$ mass distribution for the decay
$D_s^+\rightarrow K^+K^-\pi^+$ (a) and  $D_s^+\rightarrow
K^+K^-\pi^+\pi^0$ (b) as observed by BaBar~\cite{BaBar2317}.
Right: $D_s\pi^0$ mass distribution (a) and  mass
difference $\Delta M(D_s\pi^0)=M(D_s\pi^0)-M(D_s)$ (b)
as measured by Cleo~\cite{CLEO}.}
  \label{Dsj2317}
 \end{center}
\end{figure}

In April 2003 the BaBar Collaboration reported the observation of a narrow peak  in
the $D_s\pi^0$ invariant mass distribution
obtained in the charm continuum,  with mass close to $2.32$
GeV and width consistent with the experimental resolution  \cite{BaBar2317}.
The resonance, named $D_{sJ}^*(2317)$, was observed in both the $\phi\pi^+$
and $\overline{K}^{*0}K^+$ decay modes of the $D_s^+$. The peak was also found
by reconstructing $D_s$  through
$D_s \to K^+K^-\pi^+ \pi^0$.  Fig.~\ref{Dsj2317} shows the BaBar signal for
$D_{sJ}^*(2317)$.
No evidence for  $D_{sJ}^*(2317)\rightarrow D_s\gamma, D^*_s \gamma$ and
$D_s\gamma\gamma$ was  found.

The resonance was also observed  by Belle~\cite{Belle continuo} and Cleo Collaborations ~\cite{CLEO}
(fig.~\ref{Dsj2317}), with mass reported in Table~\ref{tab:Dsj}. Even in these cases,
the measured width was compatible with the experimental resolution, thus
suggesting a smaller intrinsic width of the resonance. An
observation  was reported more recently by Focus Collaboration \cite{FOCUS2317},
 with a preliminary measurement of the mass
$M_{D_{sJ}^*(2317)}= 2323\pm2$~ MeV,  slightly above the values obtained by the other
three experiments.

\begin{figure}[t]
  \begin{center}
    \epsfig{file=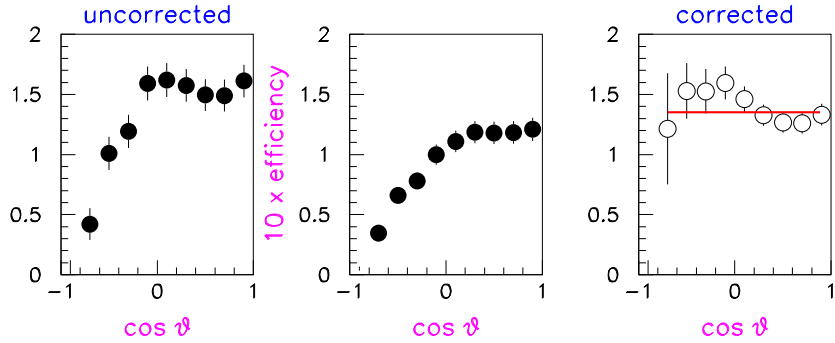,width=0.8\linewidth}
    \caption{Helicity angle distribution of the decay
$D_{sJ}^*(2317) \rightarrow D_s\pi^0$, as analyzed by BaBar \cite{Porter}.
The three panels show the
uncorrected angular distribution (left), the efficiency (center)
and the  angular distribution corrected by efficiency (right).}
    \label{babarangolare}
  \end{center}
\end{figure}

The observation of the decay $D_{sJ}^*(2317) \to D_s\pi^0$ implies for $D_{sJ}^*(2317)$
natural spin-parity.  The helicity angle distribution of
$D_s\pi^0$ obtained by BaBar (fig.~\ref{babarangolare})
is consistent with the spin 0 assignment,  even though it does not
rule out other  possibilities;  the absence of a peak in the $D_s\gamma$ final state supports the
spin-parity assignment  $J^P=0^+$. The measured mass is below the
$DK$ threshold  $s_{D^+K^0} = 2.36$ GeV.

\subsection{The meson $D_{sJ}(2460)$}
\begin{figure}
 \begin{center}
  \includegraphics[width=0.31\textwidth] {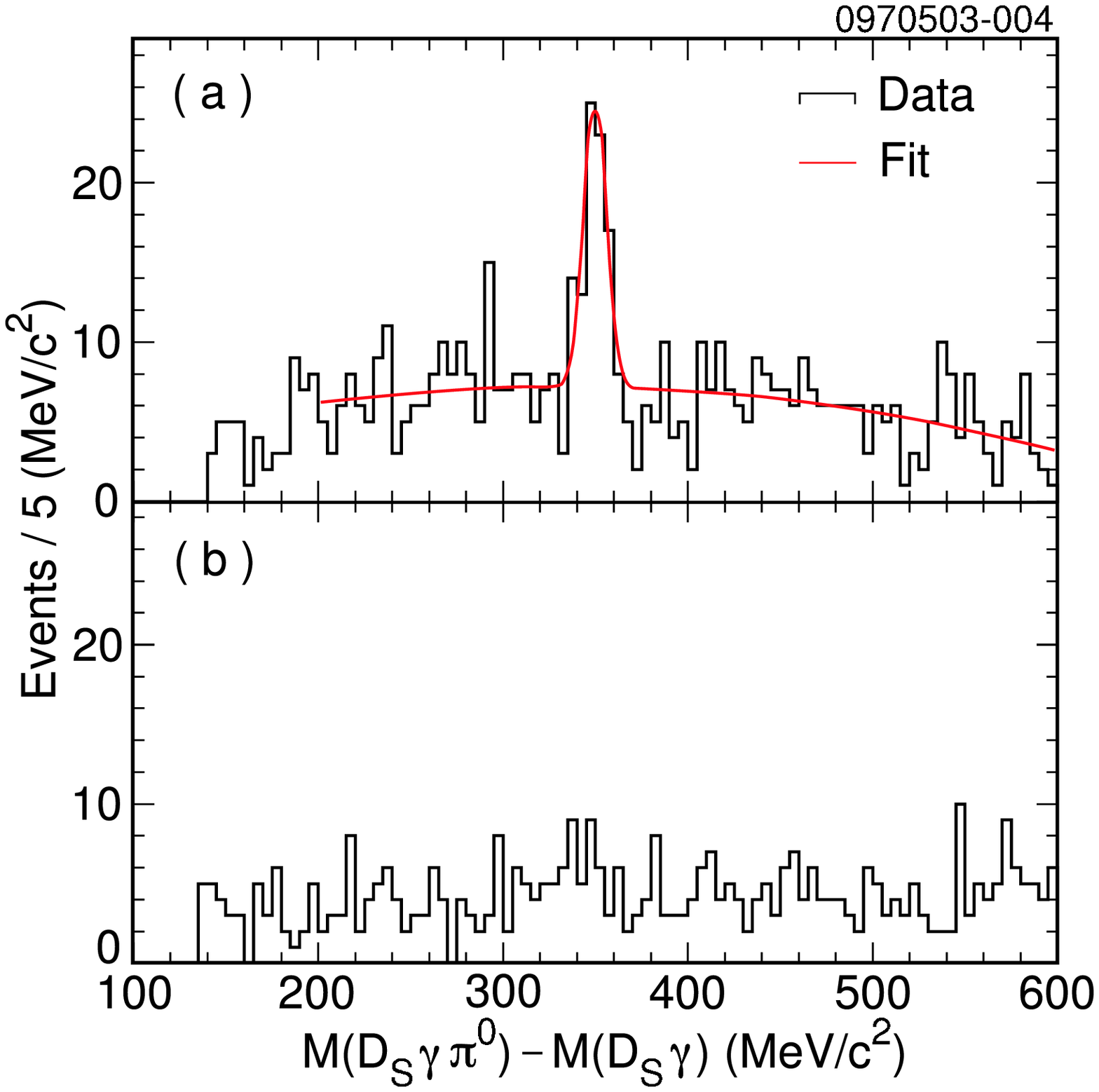}\hspace{1.5cm}
  \includegraphics[width=0.31\textwidth] {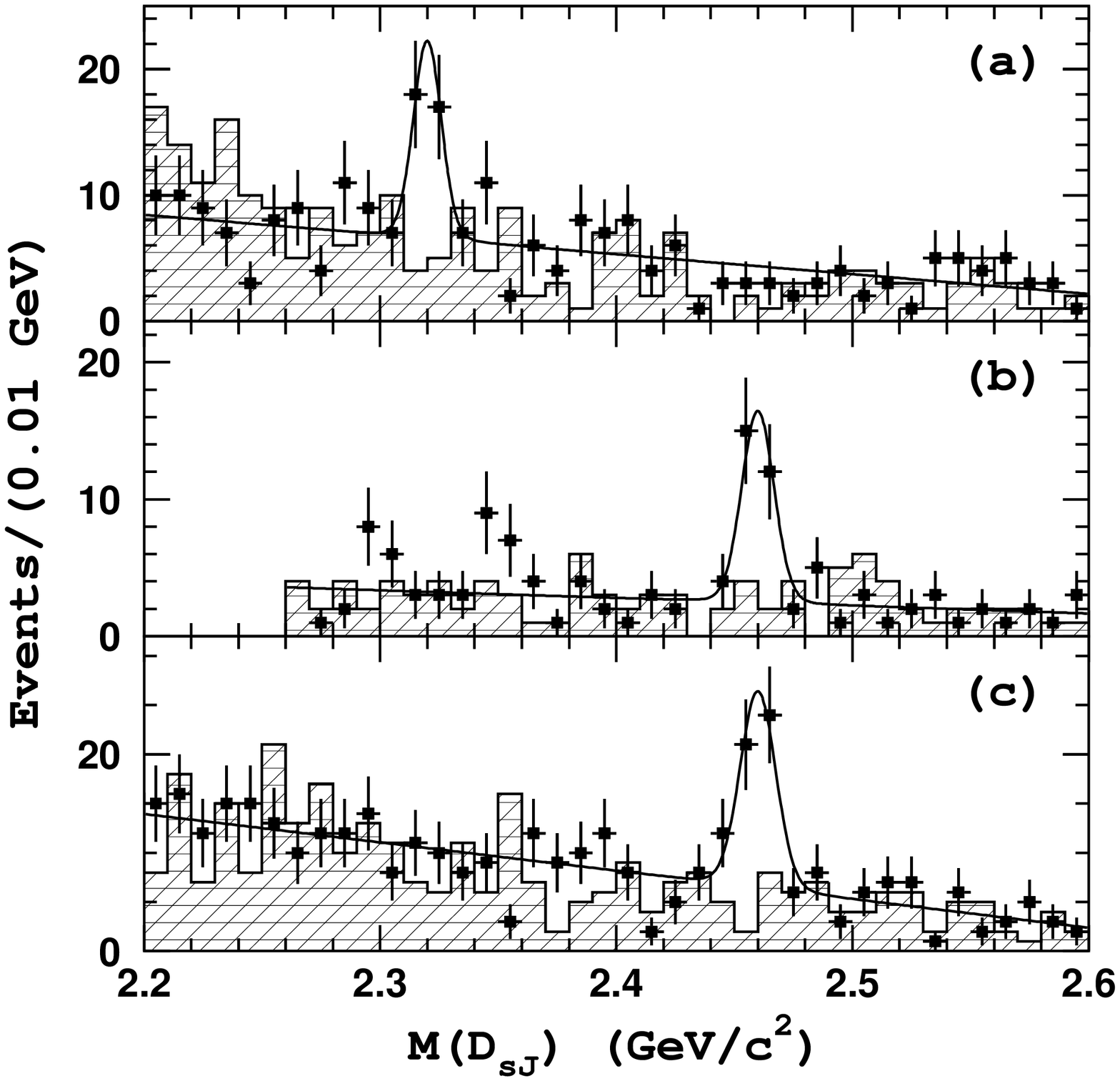}
\vspace*{-2mm}
 \caption{Left: The mass difference spectrum $\Delta
M(D_s^*\gamma\pi^0)-M(D_s\gamma)$ measured by Cleo~\cite{CLEO} (a)
for combinations where the $D_s\gamma$ system is consistent with
$D_s^*$ decay and (b) for $D_s\gamma$ combination selected from
the $D_s^*$ side band regions. Right: M($D_{sJ}$) distribution for
the $B\rightarrow\overline{D}D_{sJ}$ candidates measured by
Belle~\cite{BelledalB}: (a) $D^*_{sJ}(2317)\rightarrow D_s\pi^0$,
(b) $D_{sJ}(2460)\rightarrow D_s^*\pi^0$ and (c)
$D_{sJ}(2460)\rightarrow D_s\gamma$. }
  \label{Dsj2460}
 \end{center}
\end{figure}

Together with the $D^*_{sJ}(2317)$, Cleo Collaboration reported the
observation of
 a narrow resonance in the $D_s^*\pi^0$ system~\cite{CLEO}, with mass close to 2.46~GeV
and width consistent  with the experimental resolution.  The  peak
observed by Cleo is shown in fig.~\ref{Dsj2460}. The observation
of such a state, named D$_{sJ}(2460)$, is made difficult because
of a cross-feed ambiguity, due to the numerical relation among the
meson masses: $M(D_{sJ}(2460))-M(D_s^*)\simeq
M(D_{sJ}^*(2317))-M(D_s)\simeq 350$ MeV, which makes possible that
a $D_s^+\pi^0$ candidate is combined with a random photon such
that the $D_s^+\gamma$ combination accidentally falls in the
$D_s^{*+}$ signal region. In such a case,  $D_{sJ}^*(2317)
\rightarrow D_s^+\pi^0$ would  feed up into the $D_{sJ}(2460)
\rightarrow D_s^{*+}\pi^0$ signal region. A Monte Carlo simulation
of $D_{sJ}^*(2317)$ production and decay to $D_s^+\pi^0$ shows
that this occurs for 10\% of the reconstructed decays.

$D_{sJ}(2460)$  was also observed by Belle and BaBar,
both in the charm continuum \cite{Belle continuo,BaBar2460}, both in
$B$ decays  \cite{BelledalB,BaBardalB}, with
mass and width  reported in Table~\ref{tab:Dsj}.
In fig.~\ref{Dsj2460} we depict the Belle signal
of the radiative decay $D_{sJ}(2460)\rightarrow D_s\gamma$ reconstructed in $B$ transitions.
The measured mass in these two experiments turns out to be smaller than in Cleo
measurement, even though the results are marginally compatible: the average
of all the determinations is reported in Table~\ref{tab:Dsj}.

\begin{table}[h]
    \caption{Mass and width of the narrow resonances D$_{sJ}^*(2317)$ and
D$_{sJ}(2460)$ measured by BaBar, Belle and Cleo Collaborations. The average value for
the mass is also reported.}
 \label{tab:Dsj}
     \begin{center}
    \begin{tabular}{ccc}
    \\ \hline
      $D^*_{sJ}(2317)$ &   $D_{sJ}(2460)$ &  Collaboration\\  \hline
      \begin{tabular}{cc}
      M (GeV) & $\Gamma$ (GeV) \\  \hline
      $2317.3\pm0.4\pm0.8$ & $<10$ \\
      $2317.2\pm0.5\pm0.9$ & $<4.6$ \\
      $2318.5\pm1.2\pm1.1$ & $<$7 \\
      $2317.4\pm0.6$ &$$
      \end{tabular}
      &
       \begin{tabular}{cc}
       M (GeV) & $\Gamma$ (GeV) \\ \hline
      $2458.0\pm1.0\pm1.0$ & $<10$ \\
      $2456.5\pm1.3\pm1.3$ & $<5.5$ \\
      $2463.6\pm1.7\pm1.2$ & $<$7 \\
      $2458.8\pm1.0$ &$$
             \end{tabular}
              &
       \begin{tabular}{c}
       BaBar~\cite{BaBar2317,BaBar2460} \\
       Belle~\cite{BelledalB}             \\
       Cleo~\cite{CLEO} \\
        \end{tabular}       \\
       \hline
      \end{tabular}
  \end{center}
\end{table}

\begin{figure}[hbt]
 \begin{center}
  \includegraphics[width=0.27\textwidth] {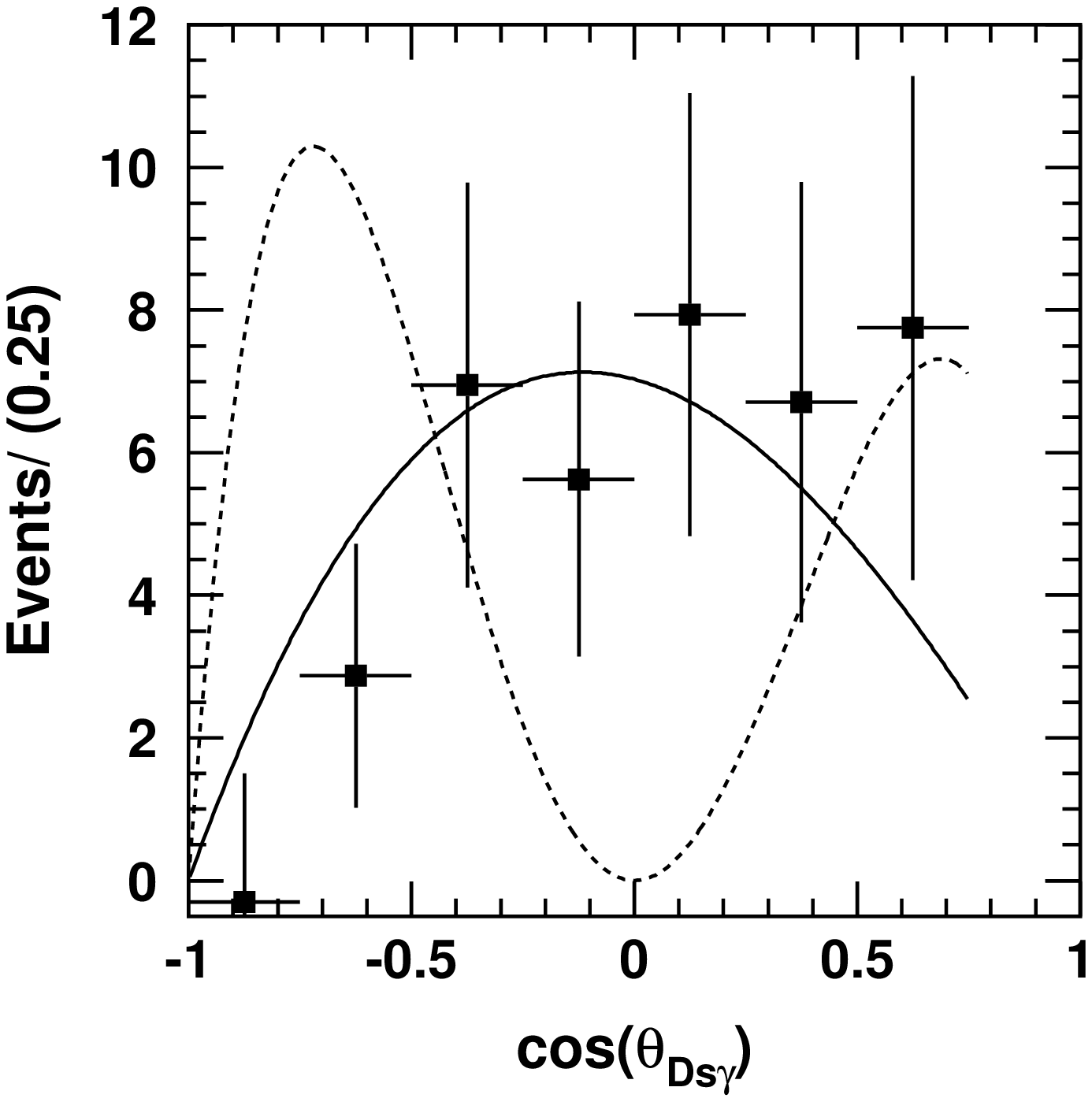}\hspace{1.2cm}
  \includegraphics[width=0.263\textwidth] {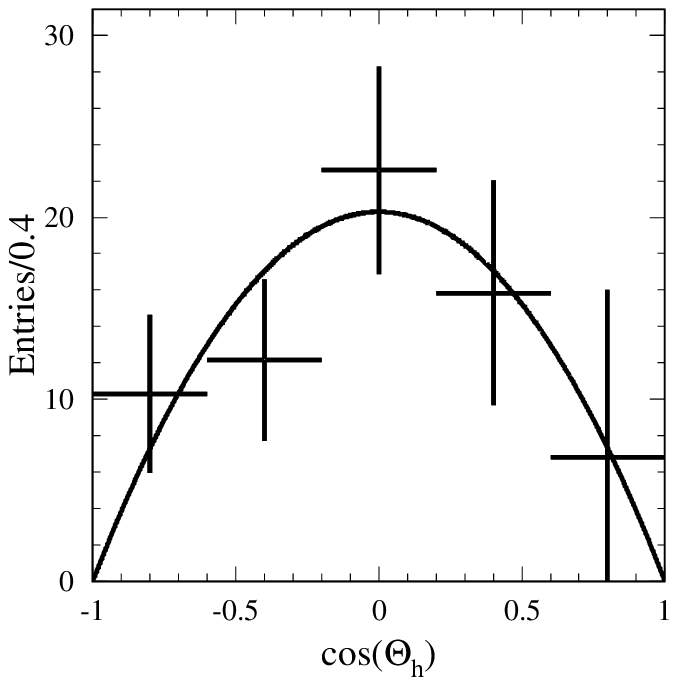} \hspace{-0.4cm}
  \includegraphics[width=0.263\textwidth] {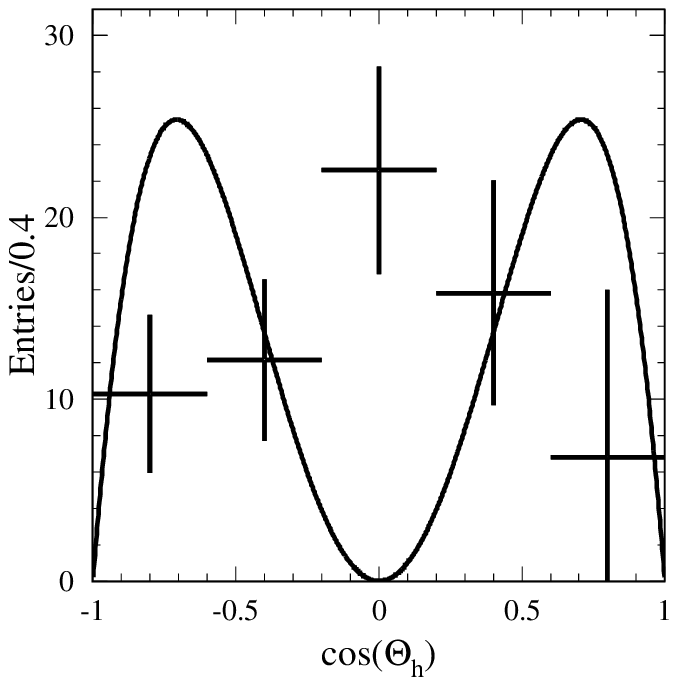}
 \caption{Helicity distribution of  D$_{sJ}(2460)\rightarrow D_s\gamma$
measured  by Belle~\cite{BelledalB} (left) and
BaBar~\cite{BaBardalB} (right).  The distributions are consistent with the assignment
$J=1$ (continuous line in the left panel, first plot in the right panel), and not with $J=2$
 (dashed line in the left panel, second plot in the right panel).}
\label{belleangolare}
 \end{center}
\end{figure}

The experimental observations are consistent with the quantum number assignment
$J^P=1^+$:  the decay D$_{sJ}(2460)\rightarrow
D_s^*\pi^0$ implies that $D_{sJ}(2460)$ has unnatural spin-parity;
the observation of  the radiative decay in $D_s\gamma$ rules out
$J=0$,  and, finally, helicity distributions measured by
Belle and BaBar in B decays are
consistent with $J=1$, as shown in  fig.~\ref{belleangolare}. The mass of  $D_{sJ}(2460)$
 is below the $D^* K$ threshold
$s_{D^{*+}K^0} = 2.51$ GeV.

The mass difference between  $D_{sJ}(2460)$ and  the other newly
observed states and the low-lying charmed mesons is reported in
Table~\ref{tab:deltam}. It is interesting to compare the hyperfine
splitting between positive and negative parity  states.
Considering the PDG values \cite{PDG}:
$M_{D^{*0}}-M_{D^0}=142.12\pm0.07 $ MeV,
$M_{D^{*+}}-M_{D^+}=140.64\pm0.10$ MeV and
$M_{D^{*}_s}-M_{D_s}=143.9\pm0.4$ MeV, one realizes  that  the
hyperfine splittings $1^+-0^+$  and $1^--0^-$ coincide in the case
of strange mesons; for non-strange mesons, the mass differences
are compatible when the Belle result for the $0^+, 1^+$ masses are
considered,  while they  disagree when the average values of
the various measurements  are considered.
\begin{table}[h]
\caption{Hyperfine splittings between positive parity mesons,  and
mass differences  between excited and low-lying $c \bar q$ and $c \bar s$ states. Belle data
in Table  \ref {tab:statilarghi} are used for the masses of the broad states.
In parentheses we also quote the results corresponding  to the   averages
in  Table \ref{tab:statilarghi}.} \label{tab:deltam}
\begin{center}
\begin{tabular}{l |  l  }
 \hline
 $\Delta M$ $(c \bar q)$  (MeV)&$\Delta M$ $(c \bar s)$  (MeV) \\  \hline
$M_{D^{'0}_1}-M_{D^{*0}_0}=119\pm26 \,\,\, (87\pm40)$&$M_{D_{sJ}(2460)} -M_{D^*_{sJ}(2317)}=141.4\pm1.2$\\ \hline
$M_{D^{'0}_1}-M_{D^{*0}}=417\pm 36\,\,\, (428\pm30)$&$M_{D_{sJ}(2460)} -M_{D_s^*}=246.4 \pm 1.2$\\
$M_{D^{*0}_0}-M_{D^{0}}=444\pm 36 \,\,\, (487 \pm 27)$&$M_{D^*_{sJ}(2317)}-M_{D_s}=348.9 \pm 0.8$\\
\hline
\end{tabular}
\end{center}
\end{table}

The measured branching fractions of two-body B decays to
$D^*_{sJ}(2317)$ or  $D_{sJ}(2460)$  are collected
in Table~\ref{tab:BR}. This is an important measurement
since, as we discuss in Section \ref{sec:decays},
hints on the nature  of the  resonances can be provided  considering   ratios of radiative
to hadronic decay rates,  either directly measured  or inferred from
data in Table~\ref{tab:BR}.

\begin{table}[hb]
\caption{Branching fractions $(10^{-3})$ of two-body B decays to
$D^*_{sJ}(2317)$ or  $D_{sJ}(2460)$, as   measured by BaBar and
Belle Collaborations. Upper limits (at  90\% C.L.) are shown in
parentheses.} \label{tab:BR}
\begin{center}
\begin{tabular}{c c c c}
 \hline
Mode & BaBar Collab.~\cite{BaBardalB} & Belle Collab.~\cite{BelledalB}& Average \\
    \hline
  $B^0\rightarrow D_{s0}^{*+}D^{-}$ $(D_{s0}^{*+}\rightarrow
D_{s}^{+} \pi^0)$  &
 $2.09 \pm 0.40\pm 0.34 ^{+0.70}_{-0.42}$  & $0.86\pm0.26{}^{+0.33}_{-0.26}$&$1.09\pm0.38$\\
  $B^0\rightarrow D_{s0}^{*+}D^{*-}$ $(D_{s0}^{*+}\rightarrow
D_{s}^{+} \pi^0)$
& $1.12 \pm 0.38\pm 0.20^{+0.37}_{-0.22}  $ & \ \hfill --- \hfill\   & \ \hfill --- \hfill\ \\
  $B^+\rightarrow D_{s0}^{*+}\overline{D}^{0}$
$(D_{s0}^{*+}\rightarrow  D_{s}^{+} \pi^0)$
& $1.28 \pm 0.37\pm 0.22 ^{+0.42}_{-0.26}$ & $0.81\pm0.24{}^{+0.30}_{-0.27}$&$0.94\pm0.32$ \\
  $B^+\rightarrow D_{s0}^{*+}\overline{D}^{*0}$
$(D_{s0}^{*+}\rightarrow  D_{s}^{+} \pi^0)$
& $1.91 \pm 0.84\pm 0.50 ^{+0.63}_{-0.38}$ & \ \hfill --- \hfill\ & \ \hfill --- \hfill\ \\
  \hline
  $B^0\rightarrow D_{s0}^{*+}D^{-}$ $(D_{s0}^{*+}\rightarrow
D_{s}^{*+} \gamma)$  &
 \ \hfill --- \hfill\  & $0.27^{+0.29}_{-0.22}(<0.95)$& \ \hfill --- \hfill\ \\
  $B^+\rightarrow D_{s0}^{*+}\overline{D}^{0}$
$(D_{s0}^{*+}\rightarrow  D_{s}^{*+} \gamma)$ &  \ \hfill ---
\hfill\ & $0.25^{+0.21}_{-0.16}(<0.76)$ & \ \hfill --- \hfill\ \\ \hline

 $B^0\rightarrow D_{s1}^{\prime +}D^{-}$ $(D_{s1}^{\prime+}\rightarrow
D_{s}^{*+} \pi^0)$
& $1.71 \pm 0.72\pm 0.27 ^{+0.57}_{-0.35}$ & $2.27\pm0.68{}^{+0.73}_{-0.62}$&$1.98\pm0.69$\\
 $B^0\rightarrow D_{s1}^{\prime+}D^{*-}$ $(D_{s1}^{\prime+}\rightarrow
D_{s}^{*+} \pi^0)$
& $5.89 \pm 1.24 \pm 1.16^{+1.96}_{-1.17} $ & \ \hfill --- \hfill\ & \ \hfill --- \hfill\ \\
  $B^+\rightarrow D_{s1}^{\prime+}\overline{D}^{0}$
$(D_{s1}^{\prime+}\rightarrow  D_{s}^{*+} \pi^0)$
& $2.07 \pm 0.71\pm 0.45 ^{+0.69}_{-0.41}$ & $1.19\pm0.36{}^{+0.61}_{-0.49}$&$1.45\pm0.59$\\
  $B^+\rightarrow D_{s1}^{\prime+}\overline{D}^{*0}$
$(D_{s1}^{\prime+}\rightarrow  D_{s}^{*+} \pi^0)$
& $7.30 \pm 1.68 \pm 1.68^{+2.40}_{-1.43}$ & \ \hfill --- \hfill\ & \ \hfill --- \hfill\ \\
  \hline
  $B^0\rightarrow D_{s1}^{\prime+}D^{-}$ $(D_{s1}^{\prime+}\rightarrow
D_{s}^{+} \gamma)$
& $0.92 \pm 0.24\pm 0.11 ^{+0.30}_{-0.19}$ & $0.82\pm0.25{}^{+0.22}_{-0.19}$&$0.86\pm0.25$\\
  $B^0\rightarrow D_{s1}^{\prime+}D^{*-}$ $(D_{s1}^{\prime+}\rightarrow
D_{s}^{+} \gamma)$
& $2.60 \pm 0.39   \pm 0.34^{+0.86}_{-0.52}$ & \ \hfill --- \hfill\  & \ \hfill --- \hfill\ \\
  $B^+\rightarrow D_{s1}^{\prime+}\overline{D}^{0}$
$(D_{s1}^{\prime+}\rightarrow  D_{s}^{+} \gamma)$
& $0.80 \pm 0.21\pm 0.12 ^{+0.26}_{-0.16}$ & $0.56\pm0.17{}^{+0.16}_{-0.15}$&$0.63\pm0.19$\\
  $B^+\rightarrow D_{s1}^{\prime+}\overline{D}^{*0}$
$(D_{s1}^{\prime+}\rightarrow  D_{s}^{+} \gamma)$
& $2.26 \pm 0.47\pm 0.43 ^{+0.74}_{-0.44}$ & \ \hfill --- \hfill\ &  \ \hfill --- \hfill\  \\
  \hline
  $B^0\rightarrow D_{s1}^{\prime+}D^{-}$ $(D_{s1}^{\prime+}\rightarrow
D_{s}^{*+} \gamma)$
& \ \hfill --- \hfill\ & $0.13^{+0.20}_{-0.14}(<0.6)$& \ \hfill --- \hfill\  \\
  $B^+\rightarrow D_{s1}^{\prime+}\overline{D}^{0}$
$(D_{s1}^{\prime+}\rightarrow  D_{s}^{*+} \gamma)$ & \ \hfill ---
\hfill\ & $0.31^{+0.27}_{-0.23}(<0.98)$& \ \hfill --- \hfill\ \\ \hline
\end{tabular}
\end{center}
\end{table}

\section{Analyses:  can the masses of $(0^+,1^+)$ $c \bar s$ ($c \bar q$)
 mesons be reliably computed?} \label{sec:masses}

\subsection{Quark models}

Quark model estimates of the masses of $p$-wave $c{\bar s}$  ($c{\bar q}$)  states
were  of course available   before April 2003,
see   Table \ref{table-masses} (A). Since
 mixing between the two $1^+$ states has been  in general  accounted,
the two axial-vector states are reported  in Table
\ref{table-masses} (A) as the   lightest  and the heaviest  of the
mass eigenstates.

\begin{table}[h]
\caption{Several pre-2003 (A) and post-2003 (B) quark model
determinations of  the masses of $p$-wave $c{\bar s}$ ($c{\bar
q}$)  mesons.} \label{table-masses}
\begin{center}
\begin{tabular}{  c  c c  c  c }
&&(A)&&\\
\hline
$D^*_{s0} (D^*_0) $ (GeV)&  $1^+_L \, c\bar s \, (c\bar q) $ (GeV)&  $1^+_H \, c\bar s \, (c\bar q) $ (GeV) &   $D^*_{s2} (D^*_2) $ GeV)& Ref.\\ \hline
2.48 (2.40)  & 2.55 (2.46) & 2.55 (2.47) & 2.59 (2.50) &\cite{Godfrey:wj} \\
2.38 (2.27)  & 2.51 (2.40) & 2.52 (2.41) & 2.58 (2.46)& \cite{Zeng:1994vj} \\
2.388  (2.279) & 2.521 (2.407)  & 2.536 (2.421)  & 2.573 (2.465) &\cite{Gupta:1994mw} \\
2.508 (2.438)  & 2.515 (2.414) & 2.569 (2.501) & 2.560 (2.459)  & \cite{Ebert:1997nk} \\
2.455 (2.341) & 2.502 (2.389) & 2.522 (2.407) & 2.586 (2.477) &\cite{Lahde:1999ih} \\
2.487 (2.377)  & 2.605 (2.490)  & 2.535 (2.417)  & 2.581 (2.460) &\cite{DiPierro:2001uu} \\
  \hline
\vspace*{0.7cm}
\end{tabular}
\begin{tabular}{  c  c   c }
&(B)&\\
\hline
$D^*_{s0} (D^*_0) $  (GeV)&  $D^\prime_{s1} (D^\prime_1)$ (GeV)                                       & Ref.\\ \hline
                                  &$2.408$  ($2.400$)                              &\cite{Cahn:2003cw}\\
  $2.357$  ($2.20$)&$2.453$  ($2.35$)&\cite{:2003dp}\\
                                  &$2.442 \pm 0.033$                                &\cite{Deandrea:2003gb}\\
 $2.288$ ($2.2$) & $2.465$ ($2.383$)                                     &\cite{Sadzikowski:2003jy}\\
 $2.446$                  &$2.515$                                                   &\cite{Lucha:2003gs}\\
 \hline
\end{tabular}

\end{center}
\end{table}

Considering  Table \ref{table-masses} (A)   one  realizes that the
mass of the scalar  $c \bar s$  was always predicted  above the
$DK$ threshold of $2.36$ GeV; therefore  such
state was expected to be massive enough to decay through isospin
conserving modes, with  a broad width.  For the axial vector state,
a few determinations also predicted mass values  close to the
$D^*K$ threshold $s_{D^{*+}K^0} =2.51$ GeV,  with  the
possibility  of having a  narrower state \cite{Charles:1998vf}.
Moreover, the non-strange mesons were always predicted to be lighter than the strange
one, with typical mass splitting of $70-100$ MeV in case of $0^+$ mesons.

The discrepancy essentially between the observed  mass and width
of $D^*_{sJ}(2317)$ and the expectation  has prompted a number of
analyses  aimed either at refining the results   to corroborate
the $c{\bar s}$ interpretation,  or at providing interpretations
in different frameworks.

Results of new mass determinations  (or elaborations of previous analyses)
are presented in Table \ref{table-masses} (B).
For example,  in a quark model with
short-distance Coulomb and   long-distance  scalar potential, with
spin-spin,  spin-orbit and  tensor terms,  using  as an input
the experimental values
$M_{D^*_2}=2.459$ GeV, $M_{D_1}=2.422$ GeV
and  $M_{D^*_0}=2.290$ GeV,  a prediction for
$M_{D_1^\prime}$ has been obtained for the non-strange
axial resonance \cite{Cahn:2003cw}.
In the $c \bar s$  case,    using
$M_{D^*_{s2}}=2.572$ GeV,
$M_{D_{s1}}=2.536$ GeV , together with
$M_{D^*_{s0}}=2.317$ GeV,  and choosing between two
possible solutions the one corresponding to a narrow
$D_{sJ}(2536)$,  a mass close to that of  $D_{sJ}(2460)$ is obtained.
Other   determinations are based on
 the Cornell potential   \cite{:2003dp},  but  the masses of
 non-strange resonances are not reproduced.
Using a further  version of the  constituent  quark model,  the spin averaged mass  of
$(D^*_{s0},D_{s1}^\prime)$ has been  derived:
$\displaystyle{M_{D^*_{s0}}+3M_{D_{s1}^\prime} \over 4}=2411 \pm 25$
 MeV   \cite{Deandrea:2003gb},  which gives the mass of $D_{s1}^\prime$  if
 the $0^+$ state is identified with  $D^*_{sJ}(2317)$.
The old fashioned MIT bag model has
 been reconsidered    \cite{Sadzikowski:2003jy}.
In general,  adjustments  of   input parameters produce
{\it a posteriori}  results in better
agreement with   observation. An exception is  a model where,
using Coulomb+linear potential and considering lowest order relativistic
corrections, the two newly observed $c \bar s$ states do not fit with the
theoretical results, thus suggesting
a  different interpretation \cite{Lucha:2003gs}.

However, even in  cases where updated results  more favourably can be compared to  data,
it is unclear   why
previous determinations resulted to be incorrect,
and what is  the new physics information that must be
encoded in models to reproduce the experimental measurements.

\subsection{Masses of  $(0^+,1^+)$ $c \bar s$
 ($c \bar q$) mesons:  non perturbative methods at work}

Since quark models suffer of not  having
 a direct relation with the QCD structure of  strong interactions,
one could look at more fundamental approaches, namely  lattice QCD and
QCD sum rules.

Lattice results for the heavy meson spectrum  are quoted in
\cite{Bali:2003jv}.  Particularly interesting are the  determinations
of the   mass difference between  the doublets  $s_\ell^P={1\over 2}^+$ and ${1\over 2}^-$,
obtained in the static limit,  either in  quenched approximation \cite{Michael:1998sg}
or for $n_f=2$ \cite{Bali:2000vr}. In  quenched approximation,
finite charm quark mass
effects were also estimated, either using  NRQCD (to order $1/m^2$
\cite{Hein:2000qu} and  $1/m^3$ \cite{Lewis:2000sv}) or
relativistic charm quarks \cite{Boyle:1997rk}, see
Table \ref{table:bali}.
%
\begin{table}
\caption{Lattice results for the mass difference (in MeV) between the doublets
$s_\ell^P={1\over 2}^+$ and ${1\over 2}^-$. The uncertainty is quoted in parentheses.}
\label{table:bali}
 \begin{center}
\begin{tabular}{ccccc}
\hline
Meson&\multicolumn{3}{c}{$n_f=0$}&$n_f=2$\\
&static&NRQCD&relativ.&static\\ \hline
$c\overline{s}$&384 (50)&465(50)&495(25)&468(43)\\
$c\overline{d}$&299(114)& ---   &465(35)&472(85) \\
\hline
\end{tabular}
\end{center}
\end{table}
%
Relativistic effects increase the mass splitting with respect to
the static case.  If the effect persists  in  unquenched
determinations, one  would obtain  a mass splitting  between
positive and negative parity doublets $\Delta M \simeq 600$ MeV,
giving $M_{D^*_{s0}}=2.57 \pm 0.11 $ GeV. On this  basis,  it  has
been  argued that lattice predictions are inconsistent with the
simple $q{\bar q}$ interpretation for  $D^*_{sJ}(2317)$
\cite{Bali:2003jv}. In a different   analysis, the continuum limit
in quenched QCD  is considered \cite{Dougall:2003hv}, and the mass
splittings $1^+-1^-$ and $0^+-0^-$ turn out to be equal:
$M_{D^*_{s0}}-M_{D_s}=389 \pm 47 $ MeV ($M_{D^*_{s0}}-M_{D_s}=435
\pm 57 $ MeV using different input parameters), compatible with
the experimental values. Now the conclusion is  that there is no
discrepancy between lattice predictions and experiment,  and
 no need to invoke other interpretations for
$D^*_{sJ}(2317)$ and $D_{sJ}(2460)$.

The absence of  definite consensus
about  $D^*_{sJ}(2317)$ can be interpreted as a difficulty in reliably
controlling the uncertainties in mass determinations by lattice QCD, at the level requested
by the available spectroscopy, and
an improvement in the systematics  seems to be necessary
to eventually reconcile the different conclusions.

As for  QCD sum rules,  the binding energies  in eq.(\ref{massformula})
 were estimated:
${\bar \Lambda}=0.5 \pm 0.1$ GeV \cite{Neubert:1991sp} and  ${\bar
\Lambda}^+=1.0 \pm 0.1$ GeV \cite{Colangelo:1998ga} for negative
and positive parity low-lying doublets, respectively. From
$M_{0^+}-M_{0^-}\simeq {\bar \Lambda}^+ - {\bar \Lambda}+{\cal
O}\left( \displaystyle{1 \over m_Q} \right)$, neglecting
$\displaystyle{1 \over m_Q}$ terms,  one obtains:
$M_{D^*_{s0}}-M_{D_s}= 500 \pm 140$ MeV (versus the experimental
data in Table \ref{tab:deltam}). A new   QCD sum rule
analysis has provided  ${\bar \Lambda}^+=0.86 \pm 0.1$ GeV  and,
including $\displaystyle{1 \over m_Q}$ corrections,
$M_{D^*_{s0}}=2.42 \pm 0.13$ GeV  \cite{Dai:2003yg}.
 Therefore, QCD sum  results seem compatible with the identification of
$D^*_{sJ}(2317)$ with the scalar $c{\bar s}$ meson,  even though
the accuracy of the mass determinations is not  high.    As
discussed in Section \ref{sec:decays},   the calculation of the
strong coupling  governing the two-body decays of  $D^*_{s0}$,
$D_{s1}^\prime$ and of their non-strange partners is useful  for
the interpretation.

\subsection{The chiral partners  $(0^-,1^-)$-$(0^+,1^+)$}

It was  suggested that a consistent implementation of chiral symmetry breaking
requires  chiral partners of pseudoscalar and
vector states \cite{Nowak:1992um}, an   idea  reconsidered
in refs. \cite{Bardeen:2003kt,Nowak:2003ra}.
Heavy-light systems should appear as
parity-doubled, i.e. in pairs differing for parity and
transforming according to a linear representation of  chiral
symmetry. In particular, the doublet composed by the states having
$J^P=(0^+,1^+)$ can be considered as the chiral partner of that
with $J^P=(0^-,1^-)$  \cite{Bardeen:2003kt}.

It is possible to build an effective lagrangian for those two
doublets and their interactions with light pseudoscalar mesons,
based on both heavy quark  and chiral symmetries. A consequence is
that the   coupling $g_\pi$ governing the $(0^+,1^+) \to (0^-,1^-)
\,\, P$ transitions, $P$ being a generic light pseudoscalar meson,
obeys a  Goldberger-Treiman relation: \be g_\pi={\Delta M \over
f_\pi} \label{gt} \ee with $\Delta M=M(0^+)-M(0^-)$. This relation
is  analogous to the one  involving the pion-nucleon coupling
constant, $\displaystyle g_{NN\pi}={m_N \over f_\pi}$. Since in
heavy-light system there is a single light constituent quark,
while the nucleon contains three,  one could expect $g_\pi \simeq
\displaystyle{g_{NN\pi} \over 3}$ and $\Delta M \simeq
\displaystyle{m_N \over 3}$. Identifying the BaBar state with the
$0^+$ $c{\bar s}$ state, one has:  $\Delta M \simeq 349 $ MeV and
$g_\pi \simeq 3.73$.

In this scheme, by suitably choosing  the parameters entering in
the terms of the lagrangian responsible of the  hyperfine
splitting in each multiplet,  one can obtain $M_{D^*_{s0}}-M_{D_s}
=M_{D_{s1}^\prime}-M_{D^*_s}$,  in agreement with the observation.
One would also obtain the relation $M_{D^*_0}-M_D
=M_{D^\prime_1}-M_{D^*}$, but the experimental results in Table
\ref{tab:deltam} are only marginally compatible with it.

It is worth mentioning  that,  since  chiral partners are split by dynamically
generated  quark mass, they could give information on the
chiral symmetry property of the medium in which they are observed.
In particular, the mass splitting between chiral partners is expected to vanish  in hot
matter  when   the chiral phase transition is approached  \cite{Nowak:2003ra}.

A test of this picture relies on computing the decay rates
of $D^*_{sJ}(2317)$ and $D_{sJ}(2460)$, as  discussed in
Section \ref{sec:decays}.

\subsection{Unitarized chiral models}

A  different approach to interpret
the new $c \bar s$ states is based on  the investigation of the
singularities in the s-wave meson-meson scattering amplitude.
An extension to the charm sector of a unitarized quark model   applied to
light scalar mesons is analyzed in  \cite{vanBeveren:2003kd}. The generalization is obtained
replacing one of the effective quark mass parameters used in  light meson systems
by the mass of the charm quark.
Including the coupling to the OZI allowed DK channel, a
scalar  meson  is found with  mass $2.28$ GeV.
Analogously, in s-wave $D \pi$ amplitude  a scalar state with mass
$2.030$ GeV is found. Conventional  $c{\bar q}$ states are found with
higher mass: $M_{D^*_0} \simeq 2.64$ GeV  and $M_{D^*_{s0}} \simeq 2.79$ GeV, both
with $\Gamma \simeq 200$ MeV \cite{vanBeveren:2003kd}.

By a similar approach, heavy-light $J^P=0^+,1^+$ mesons have been
studied using a chiral SU(3) lagrangian involving heavy-light
$0^-,1^-$ fields transforming non linearly under the chiral SU(3)
group.  Charmed mesons with $J^P=0^+,1^+$ forming antitriplet and
sextet representations of the SU(3) group are predicted; on the
contrary, the linear realization of the chiral symmetry leads to
anti-triplet states only \cite{Kolomeitsev:2003ac}. The masses of
the states to be identified with $D^*_{sJ}(2317)$ and
$D_{sJ}(2460)$ turn out to be 2303 MeV and 2552 MeV, respectively,
and the existence of several new mesons is  predicted, the
experimental evidence of which is   missing. Including
subleading terms in the chiral expansion and  adjusting three new
input  parameters to reproduce the observed spectrum,  the  values
2352 MeV and 2416 MeV for the $0^+$ and  the $1^+$ state are
obtained   together with the prediction of a scalar $I=1/2$, $S=0$
state with mass 2389 MeV
 \cite{Hofmann:2003je}.

 However,  no evidence has been collected, so far, of the  new  predicted states  enriching the open charm  spectroscopy.

\subsection{Are $D^*_{sJ}(2317)$,  $D_{sJ}(2460)$ unconventional states?}

It has been  also considered the possibility of a sizeable
four-quark component in $D^*_{sJ}(2317)$  and $D_{sJ}(2460)$.
Four-quark states could be baryonum-like or molecular-like, if
they result from bound states of quarks or of hadrons,
respectively, and examples  of the second kind of states are the
often discussed  $f_0(980)$ and $a_0(980)$   when interpreted as
$K {\overline K}$ molecules.

A possible  baryonium structure for $D^*_{sJ}(2317)$ is
$c{\bar s}q{\bar q}$ with $I=0$ (q=u,d); in that case  the
 observed transition to $D_s^+ \pi^0$ would be  isospin
 violating, explaining the observed narrowness. On the other
 hand,  $c{\bar s}q{\bar q}$ with $I=1$, predicted with nearby mass,  would be broad
  \cite{Terasaki:2003qa}.

In a  molecular interpretation,  $D^*_{sJ}(2317)$ could be viewed
as a $DK$ molecule, an interpretation supported by  the mass very
close to the DK threshold
\cite{Barnes:2003dj,Lipkin:2003zk,Bicudo:2004dx}. Although the
preferred assignment for a DK molecule would be $I=0$, it is also
possible that a mixing occurs with a $I=1,I_z=0$ molecule,
analogously to what is supposed for $f_0(980)$ and $a_0(980)$.
However, the existence of such a state would imply  isospin
partners in the $D_s^+ \pi^\pm$ invariant mass distributions: CDF
Collaboration has looked for such states without finding any
evidence \cite{Porter}. The corresponding interpretation for
$D_{sJ}(2460)$,  would be a $D^*K$ molecule.

The mechanism for producing a molecular state could be a strong
flavour-singlet attraction between a pion and a $c{\bar s}$ meson,
leading to the capture of the pion by the latter
\cite{Szczepaniak:2003vy}. Numerical estimates are in favour
 of  a molecular state  with  mass close to 2317 MeV, and of
 two other scalar resonances
($D^*_0,D^*_{s0}$) with masses and widths of:
$M_{D^*_0}=2.15-2.30$ GeV, $\Gamma(D^*_0)=7-24$ MeV;
$M_{D^*_{s0}}=2.44-2.55$ GeV, $\Gamma(D^*_{s0})=17-42$ MeV, respectively,
states that still need to be experimentally confirmed.

 It is interesting to mention that a measurement of the meson elastic form
factor could, at least in principle, allow to distinguish a two-quark state
from a four-quark state, due to the different asymptotic behavior
in the space-like momentum transfer dictated by  QCD counting
rules,   namely $1/Q^2$ for $q{\bar q}$ versus
 $1/Q^6$ in the four-quark picture.  However, the practical feasibility of such a measurement
 is difficult to assess.

It has also been  considered  the possibility that a
mixing occurs between a $c{\bar s}$ state and a four-quark state,
resulting in two mesons, one of which has mass below the $DK$
threshold \cite{Browder:2003fk}.
For the masses before the mixing, it is  assumed
 a value above the $DK $ threshold for the four-quark state,
and  of 2.48 GeV for the $c{\bar s}$ state. For several values of
the parameters $({\tilde m}_0,\theta)$=(mass of the four quark
state, mixing angle)  one of the two mixed states turns out of
mass $\simeq $ 2317 MeV. For example, for ${\tilde m}_0$ just
above the $DK$ threshold and $\theta=28.8^o$, a mass of 2319.4 MeV
is obtained for the lower mixed state. The four-quark states could
decay in doubly charged final states, as $D^+K^+$. In such
scenario the radiative transition $c {\bar s} q {\bar q} \to D_s^*
\gamma$ would be suppressed by $\sin \theta$, explaining the non
observation of such a decay mode. For the $1^+$ state the mixing
between the $c{\bar s}$ state and a  $D^*K$ state should be
analogously considered.

Finally, the possibility that the $D^*_{sJ}(2317)$ is an exotic
particle has been discussed. Strong decays of a non exotic ${\bar
q}q$ state are suppressed due to the necessity of creating a
second ${\bar q}q$ pair, while an exotic state simply falls apart
into its constituent non-exotic hadrons without any suppression,
hence they   should be broad.  If $D^*_{sJ}(2317)$ and
$D_{sJ}(2460)$ were  exotic states, they would be a special case
of   narrow exotics. A suggestion  \cite{Nussinov:2003uj}
considers $D^*_{sJ}(2317)$  a superposition of three components:
\be \ket{D^*_{sJ}(2317)}= \alpha \ket{c {\bar s}}+ \beta \ket{c {\bar s}
{({\bar u}u +{\bar d}d )\over \sqrt{2}}} + \gamma \ket{ {(K^+ D^0+
K^0D^+)\over \sqrt{2}}} \,\,  \ee \noindent giving  different
contributions when probed at different length scales. At short
scales, the pure $c {\bar s}$ component would dominate, while at
an intermediate and large scales the second component  and  the
$DK$ bound state would  prevail.

To  conclude the Section, one can  remark
that a common feature of  descriptions based on a multiquark
content of the new resonances is  the requirement  of
additional states in the spectrum, with their own decays and typical widths.
The study of  the decay modes is  an important  tool to discriminate
among different descriptions, as we discuss below.

\section{Decays of  $D_0^{*}$,  $D_1^{\prime}$, and   $D^*_{sJ}(2317)$ ,  $D_{sJ}(2460)$.}
\label{sec:decays}

In order to understand the structure of the newly observed charmed resonances, in particular
the ones with strangeness, it is necessary to analyze their decays modes and their
branching fractions. The different   interpretations  are indeed constrained to  provide
predictions in agreement with the experimental observations.

Considering  the resonances as ordinary quark-antiquark states,
one can use the heavy quark theory  together with chiral symmetry
to describe low energy interactions between heavy mesons and
pseudoscalar light mesons. A  lagrangian  invariant under heavy
spin-flavour transformations and under chiral transformations for
the pseudo Goldstone K, $\pi$ and $\eta$ bosons~\cite{hqet_chir}
\begin{equation} {\mathcal L} =igTr\{\overline{H}_aH_b\gamma_{\mu}\gamma_5{\mathcal A}_{ba}^{\mu}\}+\big \{ i hTr\{S_b\gamma_{\mu}\gamma_5{\mathcal A}^{\mu}_{ba}\overline{H}_a\}+h.c. \big \} \label{lagr} \end{equation}
involves the fields
H and  S  representing  $\frac{1}{2}^-$ and  $\frac{1}{2}^+$  doublets, respectively:
\be
H_{a}=\frac{1+\not{v}}{2}\left[ P_{a\mu }\gamma ^{\mu }-P_{a}\gamma _{5}%
\right]  \, \,\,\, , \,\,\,\,
S_{a}=\frac{1+\not{v}}{2}\left[ P_{1a}^{\mu }\gamma _{\mu
}\gamma_5-P_{0a} \right]  \label{S}
\ee
where $v$ is the meson four-velocity and $a$ is a light quark
flavour index. Light meson fields are included in
Lagrangian~(\ref{lagr}) through ${\mathcal
A}^{\mu}=\frac{1}{2}(\xi^\dagger
\partial^{\mu}\xi-\xi
\partial^{\mu}\xi^\dagger)$, where
$\xi=exp(\frac{i\tilde{\pi}}{f})$ and

\begin{equation}
\widetilde{\pi }=\left(
\begin{array}{ccc}
\frac{\pi _{0}}{\sqrt{2}}+\frac{\eta }{\sqrt{6}} & \pi ^{+} & K^{+} \\
\pi ^{-} & -\frac{\pi _{0}}{\sqrt{2}}+\frac{\eta }{\sqrt{6}} & K^{0} \\
K^{-} & \overline{K}^{0} & -\frac{2}{\sqrt{6}}\eta%
\end{array}
\right)   \label{pi tilde}
\end{equation}
with $f \simeq f_\pi$. The relevant   coupling in  the strong
decays of  $s_\ell^P={1\over2}^+$  resonances is $h$.

QCD sum rule analyses, based on both the light-cone expansion,  both on the
short-distance expansion in the soft pion limit, allowed to estimate this coupling:
$h\simeq -0.6$~\cite{Colangelo:1995ph} and  the leading heavy quark mass
corrections.  Using this value, together with the meson masses in Table \ref{tab:statilarghi},
one obtains $\Gamma \left(D_0^{*0}\rightarrow D^+\pi^-\right) =260 \pm 54$ MeV
and $\Gamma \left(D_1'\rightarrow D^{*+}\pi^-\right) =160 \pm
25$ MeV. If the modes with one pion essentially saturate the decay
widths, one predicts $\Gamma (D_0^{*0})=390\pm 80$ MeV and $\Gamma
(D_1')=240\pm 40$ MeV, consistent with the measurements in Table
\ref{tab:statilarghi}.

One
can  look at the analogous predictions for  the strange states.
However, in this cases the corresponding (isospin conserving) decays cannot  occur, since the
masses of  $D^*_{sJ}(2317)$  and $D_{sJ}(2460)$ are below the $DK$ and $D^*K$ thresholds,
respectively. Therefore, one has to invoke
a mechanism inducing isospin breaking $D^*_{s0}$ and $D^\prime_{s1}$
decays with the emission of a neutral pion. The
$\eta-\pi^0$ mixing which appears in the light meson chiral
lagrangian when the  light quark masses  are different from zero:
\be {\mathcal L}_m=\frac{\tilde{\mu}f^2}{4}Tr\left[\xi m_q\xi
+\xi^\dagger m_q\xi^\dagger \right] \,\,\, ,\ee
$m_q$ being the light quark mass matrix, can provide
 such a mechanism  as in $D^*_s\to D_s \pi^0 $ transitions
 \cite{Cho:1994zu};   indeed,   it has been applied to analyze
$D^*_{sJ}(2317)\to D_s \pi^0$  and  $D_{sJ}(2460)\to D^*_s \pi^0$
\cite{Bardeen:2003kt,Godfrey:2003kg,Colangelo:2003vg,:2003dp,Azimov:2004xk}.
The model is based on the decay  chain \be D^*_{sJ}(2317)\to D_s
\eta \to D_s \pi^0 \hspace*{.6cm} , \hspace*{.6cm} D_{sJ}(2460)\to
D^*_s \eta \to D^*_s \pi^0 \ee where the  virtual $\eta$ is  mixed
to $\pi^0$.
In the heavy quark limit the  couplings  of positive and negative parity states to pions and Kaons
 are all related to the same coupling  $h$, and therefore the same value used for the analysis
of the broad mesons  can  be used.  The resulting
amplitude depends on  an  isospin violating factor, the difference
$m_u -m_d$ between  up and down quark masses:
\begin{equation}
\Gamma \left( D_{s0}^*\rightarrow D_{s}\pi ^{0}\right)
=\frac{1}{16\pi}\frac{h^{2}}{f^{2}}\frac{M_{D_{s}}}{M_{D_{s0}^*}}\left(
m_{\pi ^{0}}^{2}+\left\vert \vec{q}\right\vert ^{2}\right)
\left\vert \vec{q}\right\vert \left(
\frac{m_{u}-m_{d}}{\frac{m_{u}+m_{d}}{2}-m_{s} }\right) ^{2}
\label{Ds0}
\end{equation}
($\vec q$ is the pion three-momentum in the $D_{s0}^*$ rest
frame), which shows why the state is  narrow. The amplitude
$D'_{s1}\rightarrow D_{s}^*\pi ^{0}$ is similar. The resulting
numerical predictions, collected in Table \ref{width},  are
compatible with hadronic decay widths of a few (or several) KeV,
well below the resolution of  the experiments that have observed
the mesons. If, instead of using the computed value of the
coupling, one uses existing information on other modes,  such as
$h_c \to J/\psi \pi_0$   \cite{Godfrey:2003kg}  or $\psi(2S) \to
J/\psi \pi^0$, $\psi(2S) \to J/\psi \eta$, or  $D_0^{\pm} \to
D^\pm \pi^0$ \cite{Azimov:2004xk},  one predicts  again narrow
hadronic widths. Measurement of  hadronic widths as in Table
\ref{width} is not an easy task; however, comparison of radiative
and hadronic decays can be used to get further information.

To analyze  radiative decays,   the electric dipole matrix
elements governing  the transitions $D^*_{sJ}(2317)\to D_s ^*
\gamma$  and  $D_{sJ}(2460)\to D^{(*)}_s \gamma$  must be
determined, and  quark model \cite{Bardeen:2003kt,
Godfrey:2003kg}, VMD \cite{Colangelo:2003vg} and relations with
radiative decays of other mesons have been used.  In particular,
if Dominance of the Vector $\phi$ Meson is assumed as in
\cite{Colangelo:2003vg},  one can use the strong coupling
$g_{D^*_{s0}  D_s^* \phi}$  derived through a low energy
lagrangian formalism  with  the heavy fields  coupled to light
vector mesons \cite{Casalbuoni:1992dx}.  The results   collected
in Table \ref{width}  present the  common feature of predicting a
suppressed radiative mode of the scalar state  with respect to the
hadronic mode.  Such a  suppression  is
observed experimentally;  however, since the radiative mode is not forbidden,  observation at the  tipical
level predicted in Table \ref{width} is expected, otherwise  different interpretations
have to be invoked.  For the axial-vector state
$D_{sJ}(2460)$, the  branching fraction of the radiative decay  into $D_s \gamma$
has been measured by Belle Collaboration.
It is interesting to   compare the  results based on $q \bar q$ interpretation
with those coming from the
 view-point of   considering $D^*_{sJ}(2317)$ as
a four-quark state  \cite{Cheng:2003kg}.  A scalar four-quark
state might be lighter than a scalar $q {\bar q}$ state with
$\ell=1$ because of the absence of the orbital angular momentum
barrier.  Multiplets of the kind $c {\bar q}_1 q_2 {\bar q}_3$
with $q=u,d,s$ can be built and the isosinglet $D^*_{s0}$ can be
identified with $D^*_{sJ}(2317)$. Assuming  that
 $D^*_{s0} \to D_s \pi^0$ proceeds through $D_{sJ} \to D_s \eta (\eta^\prime)$,
followed by $\eta(\eta^\prime) - \pi^0$ mixing, invoking
 $\eta-\eta^\prime$ mixing,
 and using for the $D_{s0}^* D_s \eta$ coupling
either  the vertex $ k K \pi$ and  SU(4) symmetry, or  QCD
sum rules \cite{Colangelo:1995ph}, a prediction of a larger  hadronic width
 than in the standard interpretation is obtained.

\begin{table}[h]
\caption{Estimated width (KeV) of $D^*_{s0}$ and
$D_{s1}^\prime$, using the $c \bar s$ picture (first five columns) or a composite
picture (last two columns).
 The results for  $D_{s1}^\prime$ in
column \cite{Colangelo:2003vg} are new. The results in column
\cite{Azimov:2004xk} are obtained using $D_0^{*0}$ decay width as measured by
Belle (Focus). } \label{width}
\begin{center}
\begin{tabular}{  l  c c  c  c c | c  c} \hline
Decay mode & \cite{Bardeen:2003kt} & \cite{Godfrey:2003kg}
& \cite{Colangelo:2003vg}& \cite{:2003dp} &
\cite{Azimov:2004xk} & \cite{Cheng:2003kg} & \cite{Ishida:2003gu}
\\ \hline $D^*_{s0} \to D_s \pi^0$ & 21.5  & $\simeq 10$  & $7 \pm 1$
 & 16  & $129 \pm 43(109\pm16)$    &10-100  &
$155 \pm 70$  \\
$D^*_{s0} \to D_s^* \gamma$
& 1.74  & 1.9  & $0.85 \pm 0.05$
 & 0.2  & $\le 1.4$  &  & 21  \\
\hline $D_{s1}^\prime \to D_s^* \pi^0$ & 21.5  & $\simeq 10$& $7 \pm
1$
& 32  & $187 \pm 73(7.4\pm2.3)$ &  & $155 \pm 70$  \\
$D_{s1}^\prime \to D_s \gamma$ & 5.08  & 6.2& $3.3 \pm 0.6$
&
& $\le 5$  & & \\
$D_{s1}^\prime \to D_s^* \gamma$ & 4.66  & 5.5& 1.5  & & & & 93  \\
\hline
\end{tabular}
\end{center}
\end{table}

Experimental information concerning ratios of radiative to
hadronic decay rates can be obtained
indirectly,    using  the branching fractions
of B decays. Although such an estimate is admittedly
uncertain, due to the correlations between various measurements of
B decay rates into positive parity charmed mesons, nevertheless it
can help  to get  hints  on the role of  radiative modes of $D^*_{s0}$, $D^\prime_{s1}$
versus the hadronic ones. In Table \ref{tab:frazioni} we have
estimated ratios of branching fractions using  B
decay data  collected in  Table \ref{tab:BR}, neglecting any
correlation among   experimental
measurements. The overall comparison of measurements with
predictions seems to support the description of the charmed
resonances as ordinary $q \bar q$ mesons.

\begin{table}[h]
\caption{Decay fractions of $c \bar s$  $\frac{1}{2}^+$ states.
The values labelled by $\left( \star \right) $ are obtained from B
decay data  in Table \ref{tab:BR} and from \cite{BelledalB}. The other ones result from
Belle~\cite{Belle continuo} and Cleo~\cite{CLEO} continuum
analyses. Few predictions are also reported.} \label{tab:frazioni}
    \begin{center}
    \begin{tabular}{cccc|ccc}
\hline & Belle & BaBar & Cleo &\cite{Colangelo:2003vg}
&\cite{Bardeen:2003kt} & \cite{Godfrey:2003kg}\\ \hline
$\frac{\Gamma \left( D^*_{s0}\rightarrow D_{s}^{\ast }\gamma
\right) }{ \Gamma \left( D^*_{s0}\rightarrow D_{s}\pi ^{0}\right)
}$ &
\begin{tabular}{c} $\left(
\star \right) \ 0.29\pm 0.26$ $\left( <0.9\right) $ \\ $<0.18$
\end{tabular}
 & \ \hfill --- \hfill\ & $<0.059$ & $0.1$ &$0.08$&$0.2$\\
\hline
$\frac{\Gamma \left( D_{s1}^{\prime }\rightarrow D_{s}\gamma \right) }{
\Gamma \left( D_{s1}^{\prime }\rightarrow D_{s}^{\ast }\pi
^{0}\right) }$ &
\begin{tabular}{c} $\left(
\star \right) \ 0.38\pm 0.11\pm0.04$  \\ $0.55\pm 0.13\pm 0.08$
\end{tabular} & \begin{tabular}{c}
$\left(\star \right)   0.44\pm 0.17$  
\end{tabular}
&$<0.49$ & $0.5$&$0.24$& $0.6$\\
\hline $\frac{\Gamma \left( D_{s1}^{\prime }\rightarrow
D_{s}^{\ast }\gamma \right) }{\Gamma \left( D_{s1}^{\prime
}\rightarrow D_{s}^{\ast }\pi ^{0}\right) }$ &\begin{tabular}{c}
$\left( \star \right) \ 0.15\pm 0.11$ $\left( <0.4\right) $  \\
$<0.31$
\end{tabular} &\ \hfill --- \hfill\
&$<0.16$ & $ 0.2$ &$0.2$&$0.6$\\
\hline
\ $\frac{\Gamma \left( D_{s1}^{\prime }\rightarrow D_{s}^{\ast
}\gamma \right) }{\Gamma \left( D_{s1}^{\prime }\rightarrow
D_{s}\gamma \right) }$ & $\left(\star \right) $ $0.40\pm 0.28$ $\left( <1.1\right) $ & \ \hfill --- \hfill\ & \ \hfill --- \hfill\ & $0.4$ &$0.9$&$0.9$\\
\hline
\end{tabular}
\end{center}
\end{table}

\section{The case of beauty}\label{sec:b}

In the simple $q \bar q$ picture, predictions for the mass of the $0^+,1^+$  $b \bar s$, $b \bar q$  states
can be obtained using data in the charm sector  together with  eq.(\ref{massformula}).
They are collected in Table \ref{table:bmasses}.
The main feature of such estimates is that the $b \bar s$ mesons are  predicted
below the $BK$ and $B^*K$ thresholds (with the exception of \cite{Bali:2003jv});
consequently,  narrow resonances in $B_s \pi^0$ and $B^*_s \pi^0$
mass distributions are expected to be observed  at the hadron colliders.

\begin{table}[h]
\begin{center}
\caption{Recent predictions for the masses of $p$-wave $b{\bar s}$
($b{\bar q}$)  mesons with $s_\ell^P={1\over2}^+$. }
\label{table:bmasses}

\begin{tabular}{  c  c   c }
\hline
$B_{s0}^* \,\, (B_0^*)$  (MeV)&  $B_{s1}^\prime \,\, (B_1^\prime)$ (MeV) & Ref.\\ \hline
$5710\pm25$     &$5770\pm25$         &\cite{Deandrea:2003gb}\\
$5654$ $(5576)$   &$5716$  $(5640)$       &\cite{Sadzikowski:2003jy}\\ \hline
$5837\pm43\pm22$&                             &\cite{Bali:2003jv}\\
$5752\pm31$&$5803\pm31$             &\cite{Green:2003zz}\\ \hline
$5718\pm35$&$5765\pm35$               &\cite{Bardeen:2003kt} \\
$5721$ ($5710-5736$)&$5762$ ($5744-5761$)   &this paper \\\hline
\end{tabular}

\end{center}
\end{table}

\section{Conclusions}\label{sec:conc}
We have briefly reviewed the experimental status of  recently observed positive parity
charmed states, as well as several
theoretical analyses devoted to determine their masses and decay
rates and to interpret the measurements.
In particular, in the case of charmed-strange states, we
have described  various interpretations  proposed for their structure.

At present, we believe that there is no compelling  evidence that a non-standard
scenario, different from  simple $q \bar q$,   is required to explain the nature of
$D^*_{sJ}(2317)$ and $D_{sJ}(2460)$,   a conclusion mainly based on the
analysis of the decay modes.

Nevertheless, unanswered  questions remain, namely about the near equality of
the masses of strange and non-strange states, as well as the
difference between the mass splittings between excited and
low-lying $c{\bar q}$ and $c{\bar s}$ states
which  is not theoretically reproduced,
as shown   by   a calculation  of chiral corrections to the meson masses
 \cite{Becirevic:2004uv}.
The missing  evidence of  the radiative mode
$D^*_{sJ}(2317) \to D_s^* \gamma$ is another puzzling aspect deserving
further experimental investigations.

To gain further information on these states,  one could
at look at two-body B decays into
$D^*_{sJ}(2317)$ or  $D_{sJ}(2460)$
\cite{Suzuki:2003rs}-\cite{Cheng:2003sm}.
As a matter of fact, the modes    $B \to D_{sJ} M$ decays,  with $M=D,\pi, K$, can further
discriminate between  quark-antiquark and  multiquark
scenarios. In the  $\bar q q$ case the $B \to D_{sJ} M$ branching
ratios are expected to be of the same order of magnitude as  $B
\to D_{s}^{(*)} M$, since the $D_{sJ}$ meson decay constants are expected to be
close to those of  low-lying  $D_{s}^{(*)}$ mesons. On the other hand, in multiquark  case the decay
amplitude would  receive additional contributions from hard
scattering of all the four valence quarks, and
the branching fractions would be  suppressed by the
coupling constant and by inverse powers of heavy meson masses.
A dedicated analysis of a complete set of data, such as that reported in Table \ref{tab:BR},
is required in this context.
 A different test  is  based on
$B_s \to D_{sJ} M$ transitions
($M=\pi, \rho, K$,  etc.) that are not currently
accessible at B factories  but  can  be investigated at the hadron
machines \cite{Datta:2003re}. The ratios of branching factions involving strange and non-strange
positive parity charm resonances:
 \be
T_{D^*_{sJ}(2317)}={{\cal B}(B_s \to D^*_{sJ}(2317) M) \over {\cal B}(B_d \to D^*_0
M)} \hspace*{0.4cm} , \hspace*{0.4cm}
T_{D_{sJ}(2460)}={{\cal B}(B_s \to D_{sJ}(2460) M) \over {\cal
B}(B_d \to D_1^\prime M)} \label{tratios} \ee  are equal to one
in the heavy quark and SU(3) limit, but  deviate from that
in composite models for the two $D_{sJ}$.
Finally, looking at
decays of higher charmonium states, e.g.   $\psi(4415) \to
D^*_s D_{sJ}(2317)$,  one can   further  investigate the  charmed resonances
\cite{barnesnew}.

A detailed analysis of properties and decays of $D^*_{sJ}(2317)$
and $D_{sJ}(2460)$, together  with their non strange partners,  undoubdetly has
become a
part of the present and future Physics programme of many currently
operating experiments, for further enriching our knowledge of
flavour  Physics and, in the end,  of QCD mechanisms of confinement.

\vspace*{1cm}
\noindent {\bf Acnowledgments.}
One of us (PC)   thanks
CPhT,  \'Ecole Polytechnique, where this work was completed  and,
in particular,  Prof. T.N. Pham for discussions.
We acknowledge partial support from the EC Contract No.
HPRN-CT-2002-00311 (EURIDICE).
\newpage


\begin{thebibliography}{99}


\bibitem{zhu}
For a recent review on pentaquarks see: S-H. Zhu, hep-ph/0406204
and references therein.

\bibitem{Skwarnicki:2003wn}
Recent  results on  heavy quarkonium  are described in:
T.~Skwarnicki,
{\it Int. J. Mod. Phys.}  {\bf A19}, 1030 (2004) and in references
therein.

\bibitem{Xcc}
M.~Mattson {\it et al.}  [SELEX Collaboration],
{\it Phys. Rev. Lett.}  {\bf 89}, 112001 (2002).

\bibitem{rassegne_hqet}
For  reviews see: M. Neubert, {\it Phys. Rep.} {\bf 245}, 259
(1994);
F.~De Fazio, in "At the Frontier of Particle Physics - Handbook of
QCD", edited by M. A. Shifman, (World Scientific, 2001), page 1671.

\bibitem{PDG}
S. Eidelman  {\it et al.}  [Particle Data Group Collaboration],
{\it Phys. Lett.}  {\bf B592}, 1 (2004).


\bibitem{cleobroad}
S.~Anderson {\it et al.}  [CLEO Collaboration],
{\it Nucl. Phys.}  {\bf A663}, 647 (2000).

\bibitem{Bellelarghi}
K.~Abe {\it et al.}  [Belle Collaboration],
arXiv:hep-ex/0307021.

\bibitem{focus}
J.~M.~Link {\it et al.}  [FOCUS Collaboration],
{\it Phys. Lett.}  {\bf B586}, 11 (2004).

\bibitem{BaBar2317}
B.~Aubert {\it et al.}  [BABAR Collaboration],
{\it Phys. Rev. Lett.}  {\bf 90}, 242001 (2003).



\bibitem{Belle continuo}
Y.~Mikami {\it et al.} [Belle Collaboration], \textit{Phys. Rev. Lett.} {\bf 92}, 012002
(2004).

\bibitem{CLEO}
D.~Besson {\it et al.}  [CLEO Collaboration], \textit{Phys. Rev.} {\bf D68}, 032002 (2003).

\bibitem{FOCUS2317}
E. W. Vaandering   [FOCUS  Collaboration], arXiv:hep-ex/0406044.

\bibitem{Porter}
F.~C.~Porter [BABAR Collaboration], arXiv:hep-ex/0312019;
A.~Pompili [BABAR Collaboration],
Proceedings of QCD@Work 2003, International Workshop on QCD: Theory
and Experiment, Conversano (Italy) 2003, P. Colangelo, F. De Fazio, R.A. Fini, E. Nappi, G. Nardulli
editors,  eConf {\bf C030614} (2003) 027.

\bibitem{BaBar2460}
B.~Aubert {\it et al.}  [BABAR Collaboration],
{\it Phys. Rev.}  {\bf D69}, 031101 (2004).

\bibitem{BelledalB}
P.~Krokovny {\it et al.} [Belle  Collaboration], \textit{Phys. Rev. Lett.} {\bf 91},  262002
(2003); P.~Krokovny  [Belle Collaboration], arXiv:hep-ex/0310053.

\bibitem{BaBardalB}
G.~Calderini  [BABAR Collaboration],
arXiv:hep-ex/0405081.




\bibitem{Godfrey:wj}
S.~Godfrey and R.~Kokoski,
{\it Phys. Rev.}  {\bf D43}, 1679 (1991).

\bibitem{Zeng:1994vj}
J.~Zeng, J.~W.~Van Orden and W.~Roberts,
{\it Phys. Rev.}  {\bf D52}, 5229 (1995).


\bibitem{Gupta:1994mw}
S.~N.~Gupta and J.~M.~Johnson,
{\it Phys. Rev.}  {\bf D51}, 168 (1995).


\bibitem{Ebert:1997nk}
D.~Ebert, V.~O.~Galkin and R.~N.~Faustov,
{\it Phys. Rev.}  {\bf D57}, 5663 (1998) [Erratum {\it ibid.} {\bf
D59}, 019902 (1999)].

\bibitem{Lahde:1999ih}
T.~A.~Lahde, C.~J.~Nyfalt and D.~O.~Riska,
{\it Nucl. Phys.}  {\bf A674}, 141 (2000).

\bibitem{DiPierro:2001uu}
M.~Di Pierro and E.~Eichten,
{\it Phys. Rev.}  {\bf D64}, 114004 (2001).

\bibitem{Charles:1998vf}
J.~Charles, A.~Le Yaouanc, L.~Oliver, O.~Pene and J.~C.~Raynal,
{\it Phys. Lett.}  {\bf B425}, 375 (1998) [Erratum {\it ibid.}
{\bf B433}, 441 (1998)].


\bibitem{Cahn:2003cw}
R.~N.~Cahn and J.~D.~Jackson,
{\it Phys. Rev.}  {\bf D68}, 037502 (2003).

\bibitem{:2003dp}
Fayyazuddin and Riazuddin,
{\it Phys. Rev.}  {\bf D69}, 114008 (2004).


\bibitem{Deandrea:2003gb}
A.~Deandrea, G.~Nardulli and A.~D.~Polosa,
{\it Phys. Rev.}  {\bf D68}, 097501 (2003).

\bibitem{Sadzikowski:2003jy}
M.~Sadzikowski,
{\it Phys. Lett.}  {\bf B579}, 39 (2004).

\bibitem{Lucha:2003gs}
W.~Lucha and F.~F.~Schoberl,
{\it Mod. Phys. Lett.}  {\bf A18}, 2837 (2003).


\bibitem{Bali:2003jv}
G.~S.~Bali,
{\it Phys. Rev.}  {\bf D68}, 07150 (2003).

\bibitem{Michael:1998sg}
C.~Michael and J.~Peisa  [UKQCD Collaboration],
{\it Phys. Rev.}  {\bf D58}, 034506 (1998).

\bibitem{Bali:2000vr}
G.~S.~Bali {\it et al.}  [SESAM Collaboration],
{\it Phys. Rev.}  {\bf D62}, 054503 (2000);
B.~Bolder {\it et al.} [SESAM Collaboration],
{\it Phys. Rev.}  {\bf D63}, 074504 (2001).

\bibitem{Hein:2000qu}
J.~Hein {\it et al.},
{\it Phys. Rev.}  {\bf D62}, 074503 (2000).

\bibitem{Lewis:2000sv}
R.~Lewis and R.~M.~Woloshyn,
{\it Phys. Rev.}  {\bf D62}, 114507 (2000).

\bibitem{Boyle:1997rk}
P.~Boyle  [UKQCD Collaboration],
{\it Nucl. Phys. Proc. Suppl.}  {\bf 63}, 314 (1998);
{\it Nucl. Phys. Proc. Suppl.} {\bf 53}, 398 (1997).

\bibitem{Dougall:2003hv}
A.~Dougall, R.~D.~Kenway, C.~M.~Maynard and C.~McNeile  [UKQCD
                  Collaboration],
{\it Phys. Lett.}  {\bf B569}, 41 (2003).



\bibitem{Neubert:1991sp}
M.~Neubert,
{\it Phys. Rev.}  {\bf D45}, 2451 (1992).

\bibitem{Colangelo:1998ga}
P.~Colangelo, F.~De Fazio and N.~Paver,
{\it Phys. Rev.}  {\bf D58}, 116005 (1998).

\bibitem{Dai:2003yg}
Y.~B.~Dai, C.~S.~Huang, C.~Liu and S.~L.~Zhu,
{\it Phys. Rev.}  {\bf D68}, 114011 (2003).



\bibitem{Nowak:1992um}
M.~A.~Nowak, M.~Rho and I.~Zahed,
{\it Phys. Rev.}  {\bf D48}, 4370 (1993);
W.~A.~Bardeen and C.~T.~Hill,
%
{\it Phys. Rev.}  {\bf D49}, 409 (1994).

\bibitem{Bardeen:2003kt}
W.~A.~Bardeen, E.~J.~Eichten and C.~T.~Hill,
{\it Phys. Rev.}  {\bf D68}, 054024 (2003).

\bibitem{Nowak:2003ra}
M.~A.~Nowak, M.~Rho and I.~Zahed,
arXiv:hep-ph/0307102.




\bibitem{vanBeveren:2003kd}
E.~van Beveren and G.~Rupp,
\textit{Phys. Rev. Lett.}  {\bf 91},  012003 (2003).


\bibitem{Kolomeitsev:2003ac}
E.~E.~Kolomeitsev and M.~F.~M.~Lutz,
{\it Phys. Lett.}  {\bf B582}, 39 (2004).

\bibitem{Hofmann:2003je}
J.~Hofmann and M.~F.~M.~Lutz,
{\it Nucl. Phys.}  {\bf A733}, 142 (2004).


\bibitem{Terasaki:2003qa}
K.~Terasaki,
{\it Phys. Rev.}  {\bf D68}, 01150 (2003).
%

\bibitem{Barnes:2003dj}
T.~Barnes, F.~E.~Close and H.~J.~Lipkin,
{\it Phys. Rev.}  {\bf D68}, 054006 (2003).

\bibitem{Lipkin:2003zk}
H.~J.~Lipkin,
{\it Phys. Lett.}  {\bf B580}, 50 (2004).

\bibitem{Bicudo:2004dx}
P.~Bicudo,
arXiv:hep-ph/0401106.

\bibitem{Szczepaniak:2003vy}
A.~P.~Szczepaniak,
{\it Phys. Lett.}  {\bf B567}, 23 (2003).

\bibitem{Browder:2003fk}
T.~E.~Browder, S.~Pakvasa and A.~A.~Petrov,
{\it Phys. Lett.}  {\bf B578}, 365 (2004).

\bibitem{Nussinov:2003uj}
S.~Nussinov,
arXiv:hep-ph/0306187.




\bibitem{hqet_chir}
M.B.Wise, {\it Phys. Rev.}  {\bf D45}, R2188  (1992); G.Burdman
and J.F.Donoghue, {\it Phys. Lett.}  {\bf  B280}, 287 (1992);
P.Cho, {\it Phys. Lett.} {\bf  B285}, 145 (1992); H.-Y.Cheng,
C.-Y.Cheung, G.-L.Lin, Y.C.Lin and H.-L.Yu, {\it Phys. Rev.} {\bf
D46}, 1148 (1992). R.~Casalbuoni, A.~Deandrea, N.~Di Bartolomeo,
R.~Gatto, F.~Feruglio and G.~Nardulli,
{\it Phys. Lett.}  {\bf B292}, 371 (1992).


\bibitem{Colangelo:1995ph}
P.~Colangelo, F.~De Fazio, G.~Nardulli, N.~Di Bartolomeo and
R.~Gatto,
{\it Phys. Rev.}  {\bf D52}, 6422 (1995);
P.~Colangelo and F.~De Fazio,
{\it Eur. Phys. J.}  {\bf C4}, 503 (1998).


\bibitem{Cho:1994zu}
P.~L.~Cho and M.~B.~Wise,
{\it Phys. Rev.}  {\bf D49}, 6228 (1994).

\bibitem{Godfrey:2003kg}
S.~Godfrey,
{\it Phys. Lett.}  {\bf B568}, 254 (2003).

\bibitem{Colangelo:2003vg}
P.~Colangelo and F.~De Fazio,
{\it Phys. Lett.}  {\bf B570}, 180 (2003).

\bibitem{Azimov:2004xk}
Y.~I.~Azimov and K.~Goeke,
arXiv:hep-ph/0403082.

\bibitem{Casalbuoni:1992dx}
R.~Casalbuoni, A.~Deandrea, N.~Di Bartolomeo, R.~Gatto,
F.~Feruglio and G.~Nardulli,
{\it Phys. Lett.}  {\bf B299}, 139 (1993).

\bibitem{Cheng:2003kg}
H.~Y.~Cheng and W.~S.~Hou,
{\it Phys. Lett.}  {\bf B566}, 193 (2003).

\bibitem{Ruijgrok:cf}
T.~W.~Ruijgrok,
{\it Acta Phys. Polon.}  {\bf B34}, 6005 (2003).

\bibitem{Ishida:2003gu}
S.~Ishida, M.~Ishida, T.~Komada, T.~Maeda, M.~Oda, K.~Yamada and
I.~Yamauchi,
arXiv:hep-ph/0310061.


\bibitem{Green:2003zz}
A.~M.~Green, J.~Koponen, C.~McNeile, C.~Michael and G.~Thompson  [UKQCD
                  Collaboration],
{\it Phys. Rev.}  {\bf D69}, 094505  (2004).


\bibitem{Becirevic:2004uv}
D.~Becirevic, S.~Fajfer and S.~Prelovsek,
arXiv:hep-ph/0406296.

\bibitem{Suzuki:2003rs}
M.~Suzuki,
arXiv:hep-ph/0307118.

\bibitem{Chen:2003jp}
C.~H.~Chen and H.~N.~Li,
{\it Phys. Rev.}  {\bf D69}, 054002 (2004).

\bibitem{Cheng:2003id}
H.~Y.~Cheng,
{\it Phys. Rev.}  {\bf D68}, 094005 (2003).

\bibitem{Datta:2003re}
A.~Datta and P.~J.~O'Donnell,
{\it Phys. Lett.}  {\bf B572}, 164 (2003).

\bibitem{Huang:2004et}
M.~Q.~Huang,
{\it Phys. Rev.}  {\bf D69}, 114015 (2004).

\bibitem{Cheng:2003sm}
H.~Y.~Cheng, C.~K.~Chua and C.~W.~Hwang,
{\it Phys. Rev.}  {\bf D69}, 074025 (2004).

\bibitem{barnesnew}
T. Barnes, arXiv:hep-ph/0406327.

\end{thebibliography}
\end{document}